\documentclass[pra,10pt,aps,twocolumn,groupedaddress,%
longbibliography,showpacs]{revtex4-1}

\def\ket#1{\left|#1\right\rangle}
\def\bra#1{\left\langle#1\right|}
\def\tr{\mathop{\rm Tr}}

\makeatletter
\def\Hy@safe@activestrue{}
\makeatother
\usepackage{amsmath}
\usepackage{amstext}
\usepackage{amsfonts}
\usepackage{graphicx}
\usepackage{color}
\usepackage{float}
\usepackage{tabularx}
\usepackage{hyperref}

\graphicspath{{Figs/}}

\begin{document}

\title{Universal set of Dynamically Protected Gates for Bipartite
  Qubit Networks II: \\
  Soft Pulse Implementation of the [[5,1,3]] Quantum Error Correcting
  Code.}

\author{Amrit De and Leonid P. Pryadko} \affiliation{Department of
  Physics and Astronomy, University of California, Riverside,
  California, 92521, USA}

\date{\today}

\begin{abstract}
  We model repetitive quantum error correction (QEC) with the
  single-error-correcting five-qubit code on a network of
  individually-controlled qubits with always-on Ising couplings, using
  our previously designed universal set of quantum gates based on
  sequences of shaped decoupling pulses.  In addition to serving as
  accurate quantum gates, the sequences also provide dynamical
  decoupling (DD) of low-frequency phase noise.  The simulation
  involves integrating unitary dynamics of six qubits over the
  duration of tens of thousands of control pulses, using classical
  stochastic phase noise as a source of decoherence.  The combined
  DD/QEC protocol dramatically improves the coherence, with the QEC
  alone responsible for more than an order of magnitude infidelity
  reduction.
\end{abstract}

\maketitle
\section{INTRODUCTION}\label{sec:intro}

Quantum error correction\cite{shor-error-correct,gottesman-thesis,%
  Knill-Laflamme-1997,Terhal-RMP-2015} (QEC) makes it possible to
perform large scale quantum computations with a finite error rate per
qubit\cite{Shor-FT-1996,Steane-FT-1997,Gottesman-FT-1998,%
  Dennis-Kitaev-Landahl-Preskill-2002,Knill-FT-2003,%
  Knill-error-bound,Steane-2003}. Much like their classical
counterparts, quantum error correcting codes (QECCs) rely on adding
redundant qubits to control errors.  However, unlike, e.g., in the
classical information transmission problem, qubits remain subject to
errors all the time, in particular, during the syndrome extraction.
Thus, to achieve scalability, special fault-tolerant (FT) protocols
must be used both for QEC and for any operation with the encoded
qubits.  This increases the overhead and is one of the reasons why the
error probability thresholds to scalable quantum computation are so
small---e.g., around $1\%$ per local gate in the case of the toric and
related surface
codes\cite{kitaev-anyons,Dennis-Kitaev-Landahl-Preskill-2002,%
  Raussendorf-Harrington-2007}.  The number of qubits needed,
measurement complexity, and stringent requirements for gate speed and
fidelity are also among the reasons why an experimental demonstration
of the repetitive quantum error correction with a universal quantum
code so far remains an elusive
goal.\cite{Cory-QECC-1998,Chiaverini-2004,Pittman-linear-optics-QEC-2005,%
  Schindler-Blatt-repetitive-2011,Moussa-NMR-QECC-2011,Reed-QEC-SC-2012,%
  Martinis-44us-qubit-2013,Zhong-Wang-Martinis-Cleland-Korotkov-Wang-2014,%
  Chow-etal-Steffen-2014,Barends-etal-Martinis-2014,%
Corcoles-etal-Chow-2015,Kelly-etal-Martinis-2015}

A possible way to loosen these requirements is to combine active QEC
with one of the passive error reduction techniques depending on the
correlations in the dominant decoherence
channel\cite{Lidar-Chuang-Whaley-1998,viola-knill-lloyd-1999,%
  viola-knill-lloyd-1999B,lidar-1999,Bacon-2000,%
  Kempe-2001,Viola-2002,facchi-nakazato-2004,Lidar-review-2014}.  In
particular, dynamical decoupling (DD) is most
effective\cite{shiokawa-lidar-2004,%
  faoro-viola-2004,sengupta-pryadko-ref-2005,%
  pryadko-sengupta-kinetics-2006,kuopanportti-2008,Cywinski-2008,%
  West-Lidar-Fong-Gyure-2010} in suppressing the effects of
low-frequency (e.g., $1/f$) noise which is often the leading mechanism
for the loss of phase coherence.  Moreover, DD can be used to achieve
scalability in gate design, since pulses and sequences intended for a
large system can be constructed to a given order in the Magnus
series\cite{slichter-book} on small qubit
clusters\cite{sengupta-pryadko-ref-2005,pryadko-sengupta-kinetics-2006}.
DD is also excellent in producing accurate control for systems where
not all interactions are known as one can decouple interactions with
the given symmetry\cite{stollsteimer-mahler-2001,Tomita-2010}, and it
can be implemented to remain stable when environment has sharp
spectral features\cite{pryadko-quiroz-2007} or high-frequency
modes\cite{pryadko-quiroz-2009}, or even with substantial pulse
errors\cite{pryadko-sengupta-2008,%
  Kabytayev-Green-Khodjasteh-Biercuk-Viola-Brown-2014}.  In short, at
least in principle, using DD at the lowest level of coherence
protection could substantially reduce the required repetition rate of
the QEC cycle.  In many instances, this could make a crucial
difference, enabling the use of QEC.

Recently we made a substantial progress toward developing a combined
DD/QEC coherence protection protocol by constructing a universal set
of quantum gates based on soft-pulse DD
sequences.\cite{De-Pryadko-2013,De-Pryadko-FT-2014} The gates are
designed to work on a network of qubits with always-on Ising couplings
forming a sparse bipartite graph.  In addition to providing accurate
control, these gates also work as decoupling sequences, suppressing
the effect of low-frequency phase noise to second order in the Magnus
series.  With these gates, we demonstrated\cite{De-Pryadko-2013} the
effectiveness of repetitive QEC using single-error-detecting QECC
$[[4,2,2]]$ encoding two qubits in four by simulating full unitary
dynamics of five driven qubits in an Ising chain, using low-frequency
classical noise as the source of decoherence.

We have also studied\cite{De-Pryadko-FT-2014} the errors associated
with the gates similar to those constructed in
Ref.~\onlinecite{De-Pryadko-2013}.  In a system with always-on
pairwise qubit couplings, for any gate constructed perturbatively to a
given order $K$, only the errors forming clusters that involve up to
$K+1$ qubits are suppressed.  Large-weight clusters of correlated
errors can be suppressed exponentially when gates are executed fast
enough.  However, such a choice can only be made with a sufficiently
sparse coupling network.  Increasing the maximum degree $z$ of the
connectivity graph with pulse duration and other gates' parameters
fixed may lead to proliferation of large uncorrectable error clusters.

In our previous work\cite{De-Pryadko-2013}, we simulated a linear
Ising chain with $z=2$, an arrangement most favorable for controlling
multi-qubit correlated errors\footnote{Note that this is exactly the
  arrangement chosen for experiments in Ref.\
  \onlinecite{Barends-etal-Martinis-2014}.}.  On the other hand, the
optimal arrangement for surface codes is planar.  The corresponding
analytical bound on maximum gate duration needed for FT is
small\cite{De-Pryadko-FT-2014}.  Thus, it remains an open question
whether perturbation-theory-based gates like those constructed in
Ref.\ \onlinecite{De-Pryadko-2013} can be practical for use in
repetitive QEC.

\begin{figure}[tbp]
  \centering
  \includegraphics[width=0.8\columnwidth]{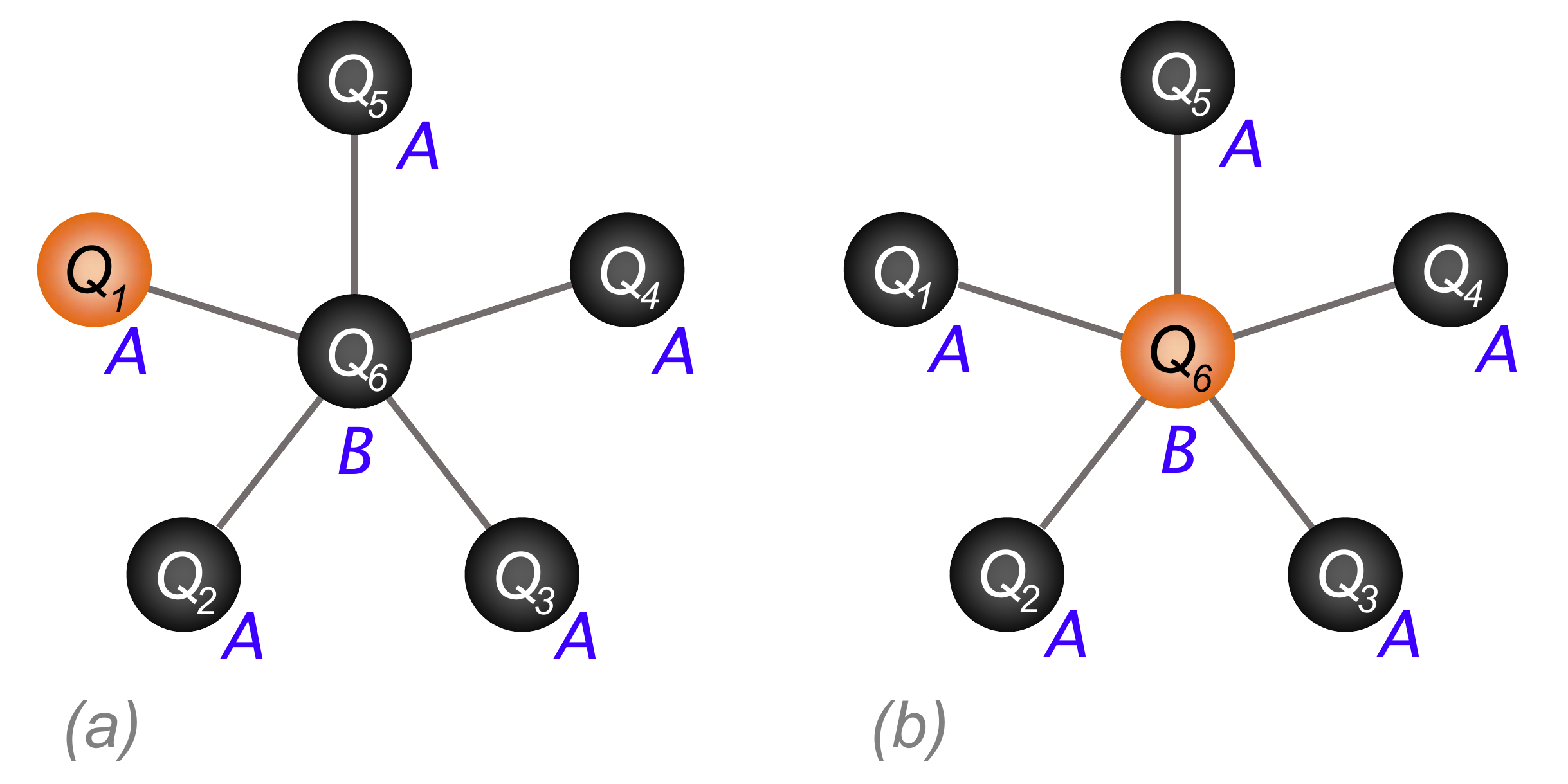}
  \caption{(Color online) Arrangement of qubits on a bipartite star
    graph for implementing the [[5,1,3]] code. This particular
    arrangement of the qubits was chosen to minimize the total number
    of gate operation during the stabilizer generator measurement
    cycle. ({\bf a}) The qubit to be encoded, $Q_6$, is initially
    placed at the center.  At the end of encoding, it is swapped with
    an ancilla qubit, initially $Q_1$.  ({\bf b}) During the
    stabilizer measurement cycle, the single ancilla qubit $Q_6$ at
    the center is used to sequentially measure all four stabilizer
    generators supported by the five qubits around it.}
  \label{fig:schematic}
\end{figure}

In this work we simulate numerically repetitive quantum error
correction using our universal gate
set\cite{De-Pryadko-2013,De-Pryadko-FT-2014} in a network with $z\ge4$
as would be needed for the surface code.  Specifically, we use a
six-qubit star graph (see Fig.~\ref{fig:schematic}) of Ising couplings
between the qubits with $z=5$, and simulate QEC with the $[[5,1,3]]$
code both in the traditional and in the error-detection (Zeno) modes.
This code can actually be seen as a variant of a surface code with
rotated periodicity
vectors.\cite{Kovalev-Dumer-Pryadko-2011,Kovalev-Pryadko-2012}  We
simulate full unitary dynamics over several error correcting cycles
(up to seventy thousand shaped pulses) with instantaneous
ancilla-based projective measurements, and classical Gaussian
correlated phase noise used as a source of decoherence.  We consider
the cases of low-frequency noise with Gaussian correlations, as well
as a bimodal noise generated as a combination of a low-frequency and a
high-frequency components.  The constructed protocols show substantial
improvement of coherence compared to the case of unprotected qubits,
including over an order of magnitude average infidelity reduction
attributable to error correction alone.

The structure of the paper is as follows.  In Sec.~\ref{sec:model}, we
give a brief overview of dynamical decoupling, of the universal gate
set used, and our implementation of the $[[5,1,3]]$ code on a star
graph.  In Sec.~\ref{sec:simulations}, we present the results of
numerical simulations.  We give Conclusions in
Sec.~\ref{sec:conclusions}.

\section{Gate and code implementation}\label{sec:model}

\subsection{Dynamical control on an Ising network}
\label{sec:dd-pulse}
The goal of dynamical control is to drive the desired unitary
evolution of a quantum system over a given time interval.  While the
details of the dynamics during the interval may differ greatly, the
net result of such an evolution can be, to some extent, insensitive to
the details of system's interactions.  For example, in the simplest
case of single-qubit dynamical decoupling, the qubit interactions are
averaged out during the period of the controlled evolution.

We consider the following general Hamiltonian
\begin{equation}\label{eq:Hmain}
  H=H_\mathrm{C}+H_0,\quad H_0 = H_{\rm S} + H_{\rm B} + H_{\rm SB},
\end{equation}
where $H_\mathrm{C}\equiv H_{\rm C}(t)$ is the time-dependent control
Hamiltonian, and the remaining Hamiltonian $H_0$ is separated into the
parts $H_\mathrm{S}$ and $H_{\rm B}$ acting on the qubit ``system''
and on the bath respectively, and the system-bath coupling Hamiltonian
$H_{\rm SB}$.

In this work, following
Refs.~\onlinecite{De-Pryadko-2013,De-Pryadko-FT-2014}, we consider a
sparse bipartite network of qubits with the Ising couplings between
nearest neighbors\footnote{Selective decoupling sequences for more
  general qubit interaction Hamiltonians have been constructed, e.g.,
  in Refs.~\onlinecite{sengupta-pryadko-ref-2005,%
    Frydrych-Marthaler-Alber-2015}.},%
\nocite{Frydrych-Marthaler-Alber-2015}
\begin{equation}\label{eq:ising-network}
  H_{\rm S}\equiv {1\over 2} \sum_{\langle i,j\rangle}
  J_{ij} \sigma^{z}_{i}\sigma^{z}_{j}, 
\end{equation} 
and decoherence due to slow dephasing of individual qubits, generally
described by the following bath and bath-coupling Hamiltonians:
\begin{equation}
  \label{eq:ising-bath}
  H_{\rm B}=\sum_i B_i,\quad H_{\rm SB}={1\over2}\sum_i A_i \sigma_i^z.
\end{equation}
Each qubit is assumed to have its own individual bath, i.e., the bath
operators $B_j$ commute with each other, and the coupling operators
$A_i$ commute with each other and all $B_j$, $j\neq i$.

The decoupling technique assumes that the control Hamiltonian $H_{\rm
  C}$ dominates the dynamics.  We implicitly assume that any large
energies have already been eliminated from the system $H_{\rm S}$ and
system-bath coupling $H_{\rm SB}$ Hamiltonians by a rotating wave
approximation (RWA).  Then, the Hamiltonian~(\ref{eq:ising-network})
can be viewed as an effective Hamiltonian for any set of interactions
as long as the transition frequencies of the neighboring qubits differ
sufficiently.  Similarly, the bath model~(\ref{eq:ising-bath}) is an
effective description of qubits operating well above the bath
frequency cut-off to eliminate direct spin flip transitions, with
dephasing caused, say, by phonon scattering.

We also assume the ability to control the qubits individually, with
the control Hamiltonian
\begin{equation}
  H_{\rm C}\equiv \sum_j H_{\rm C}^{(j)},\quad H_{\rm C}^{(j)}={1\over
    2}\sum_{\mu=x,y,z} V_{j\mu}(t)\,\sigma_j^\mu ,
  \label{eq:control-general}
\end{equation}
where the time dependent control signals $V_{j\mu}(t)$ are arbitrary,
except for some implicit limits on their amplitude and spectrum.  Our
gates\cite{De-Pryadko-2013,De-Pryadko-FT-2014} are designed as
sequences of one-dimensional pulses, with the control fields on a
given qubit applied along $x$, $y$, or $z$ axis exclusively, so that
only one of $V_x(t)$, $V_y(t)$, or $V_z(t)$ can be non-zero at any
given time.  We also imposed a restriction that no pulses be applied
simultaneously on any pair of neighboring qubits.

As a result of these assumptions, the multi-qubit unitary evolution
operator with the complete Hamiltonian~(\ref{eq:Hmain}) over the
duration of a single-pulse interval can be written as a product of
mutually commuting terms, each of them involving the controlled qubit
and various products of $\sigma^z$ Pauli operators for its
uncontrolled neighbors.\cite{De-Pryadko-FT-2014}  Each of these
operators can be computed order-by-order in the time-dependent
perturbation theory; in Ref.\ \onlinecite{De-Pryadko-FT-2014} we
carried such an expansion up to third order. In each order of the
series, the dependence on the pulse shape is encoded in terms of just
a few coefficients\cite{pryadko-quiroz-2007,%
  pryadko-sengupta-2008,pryadko-quiroz-2009,De-Pryadko-FT-2014}.  For
example, the first-order correction is expressed in terms of just two
such coefficients, the time averages of $\cos \phi(t)$ and $\sin
\phi(t)$ over the duration of the pulse, where $\phi(t)=\int_0^t dt'
V(t')$ is the time-dependent rotation angle corresponding to the given
pulse shape $V(t)$, $0\le t\le \tau_p$, and $\tau_p$ is the pulse
duration.  With generic pulse shapes, such as a Gaussian, this
produces errors that scale linearly with $\tau_p$.  Specially designed
self-refocusing pulses can be constructed to suppress this effect,
e.g., to linear or quadratic orders in powers of $\tau_p$, depending
on the shape\cite{sengupta-pryadko-ref-2005, pryadko-sengupta-2008}.
For example, to the linear order, this is done by choosing a
functional form $V(t)$ which guarantees $\langle
\cos\phi(t)\rangle=\langle \sin\phi(t)\rangle=0$.  If the pulse shape
is symmetric, $V(t)=V(\tau_p-t)$, this requires only one additional
condition on the
shape\cite{warren-herm,sengupta-pryadko-ref-2005,pryadko-sengupta-2008}.

While in a multi-qubit setting such special pulse shapes do not
eliminate all first- or second-order errors over the pulse duration,
the resulting series have fewer terms which can be subsequently dealt
with easier by properly designing the pulse sequences.

\subsection{Universal gate set}
\label{sec:dd-seqs}

With generic set of inter-qubit couplings, increasing the number of
qubits requires progressively longer sequences to decouple the
inter-qubit couplings\cite{stollsteimer-mahler-2001}.  However, when
the couplings form a bipartite graph, such a decoupling to an
arbitrary (fixed) order can be done with a single sequence of a finite
duration independent of the number of qubits in the
system\cite{De-Pryadko-2013}.  We constructed a gate set formed by an
arbitrary single-qubit rotation and an entangling controlled-phase
gate [more precisely, arbitrary-angle $ZZ$ rotation, $\exp(-i\alpha
\sigma^z_i\sigma^z_j)$ for a pair of neighboring qubits $i$ and $j$].
According to general theory, such a gate set is
universal\cite{Barenco-1995}.  These gates can be executed
simultaneously on an arbitrary set of non-neighboring qubits (pairs of
qubits), and in addition provide DD protection for every qubit.  In
particular, all the Hamiltonian terms not directly involved in the
gate are averaged out.

\textsc{Single-qubit gates}\cite{De-Pryadko-2013,De-Pryadko-FT-2014}
are based on the leading-order dynamically corrected
gates\cite{Khodjasteh-Viola-PRL-2009,Khodjasteh-Viola-PRA-2009}, in
turn based on the Eulerian path
con\-struc\-tion\cite{Viola-Knill-2003}.  The original single-qubit
DCG sequence\cite{Khodjasteh-Viola-PRL-2009,Khodjasteh-Viola-PRA-2009}
guarantees leading-order cancellation of an arbitrary bath coupling.
This is achieved by executing a sequence of identically-shaped pulses
driving the qubit through an (extended) Eulerian cycle on the Cayley
graph corresponding to the decoupling group.  Explicitly, the
single-qubit DCG
sequence\cite{Khodjasteh-Viola-PRL-2009,Khodjasteh-Viola-PRA-2009} can
be formally written as
\begin{equation}
  (\mathbf{X})(I)(\mathbf{Y})(I)(\mathbf{X})(I)(\mathbf{Y})\;
  (\mathbf{Y})(\mathbf{X})(\mathbf{Y})(\mathbf{X})\;(P),\label{eq:orig-dcg}
\end{equation}
where $(\mathbf{X})$ and $(\mathbf{Y})$ represent finite-duration
$\pi$ pulses in $X$ and $Y$ direction, $(P)$ the pulse nominally
implementing the desired single-qubit rotation, and $I$ is the
composite pulse implementing a unity operation by running a half-time
double-amplitude version of $P$ followed by an identical pulse applied
in the opposite direction.  As written, the sequence works for pulses
of arbitrary symmetric shapes, as long as these shapes remain the same
during the sequence.

To build dynamically-protected single-qubit gates on a bipartite qubit
network with always-on couplings, we separated the DCG sequence into a
part to be executed on the sublattice $A$ [$X$ pulses in the original
sequence (\ref{eq:orig-dcg})] and a part to be executed on the
sublattice $B$ ($Y$ pulses in the original sequence replaced by $X$
pulses).  Each of these are \emph{partial-group} sequences as they go
over Eulerian cycles corresponding to subgroups of the two-sublattice
decoupling group, specifically chosen to control Ising bath
coupling~(\ref{eq:ising-bath}).  As a result, the entire sequence is
only effective against dephasing, and it requires self-refocusing
pulses (see Sec.~\ref{sec:dd-pulse}) to achieve the leading-order
cancellation.

The construction allows simultaneous rotations in any set of
non-neighboring qubits (e.g., the entire sublattice $A$ or $B$ can be
rotated at once), with $P$ representing the desired rotation or zero
applied field on idle qubits.  In actual implementation we used the
stretched pulse $P$ of duration $2\tau_p$, so that the unity operation
$I$ in Eq.~(\ref{eq:orig-dcg}) is composed of two pulses of duration
$\tau_p$.  Overall, the duration of such a single-qubit gate is
$16\tau_p$.  The Hadamard gate $H$ is implemented as a product of two
rotations, with the net duration $32\tau_p$.

\begin{figure*}[htbp]
  \centering
  \includegraphics[width=0.8\textwidth]{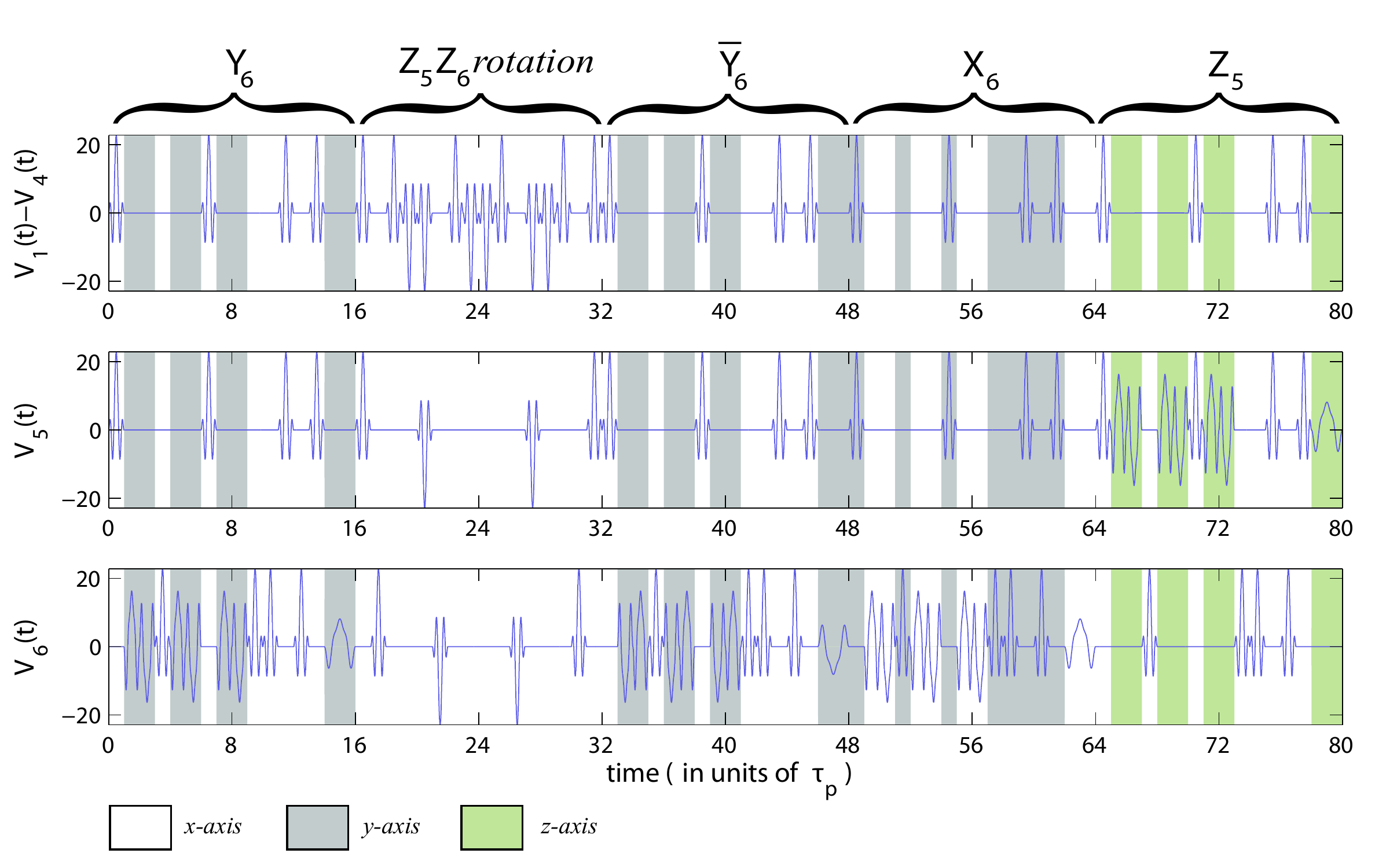}
  \caption{(Color online) Pulse sequences used to implement the CNOT
    gate between qubits $\mathcal{Q}_5$ and $\mathcal{Q}_6$ on a star
    graph.  It is a combination of four single-qubit gates
    ($0<t\le16\tau_p$, $32\tau_p<t\le 48\tau_p$, $48\tau_p<t\le
    64\tau_p$, and $64\tau_p<t\le 80\tau_p$) and a single instance of
    the $ZZ$-coupling sequence, $16\tau_p<t\le 32\tau_p$.  For better
    accuracy this latter sequence has to be repeated several times, we
    used $N_\mathrm{rep}=5$ repetitions, see
    Ref.~\onlinecite{De-Pryadko-FT-2014} for detailed discussion of
    the associated errors.  Second-order self-refocusing pulse shapes
    $Q_1(\pi)$ and $Q_1(\pi/2)$ from
    Refs.~\onlinecite{sengupta-pryadko-ref-2005,pryadko-sengupta-2008}
    are used.  The shading shows the direction of the applied pulses
    as indicated.}
  \label{fig:cnot-seqs}
\end{figure*}

Same sequences used with second-order self-refocusing
pulses\cite{sengupta-pryadko-ref-2005,pryadko-sengupta-2008} (see the
portion $t\le 16\tau_p$ on Fig.~\ref{fig:cnot-seqs}) yields
second-order cancellation of inter-qubit couplings and the bath terms
in Eq.~(\ref{eq:ising-bath}), except for terms proportional to the
commutators $[B_i,A_i]$.  These terms are readily interpreted as the
derivatives of time-dependent fields acting on the qubits.  Such terms
can also be canceled\cite{De-Pryadko-FT-2014}, e.g., using symmetrized
versions of our sequences (involving 32 pulses instead of 16), if one
uses more complicated pulse shapes akin to those developed in
Ref.~\onlinecite{Pasini-2008}.  While sequences achieving higher
cancellation orders can be readily designed using the same general
approach\cite{Khodjasteh-Viola-PRL-2009,%
  Khodjasteh-Viola-PRA-2009,%
  Khodjasteh-Lidar-Viola-2009}, the advantage of the particular
sequences we use in this work is that they are shorter.

\textsc{Two-qubit $ZZ$-rotation
  gates}\cite{De-Pryadko-2013,De-Pryadko-FT-2014} are designed using a
different approach, see Fig.~\ref{fig:cnot-seqs}.  The idea is to
selectively decouple some of the inter-qubit interactions, with the
needed rotations generated by the residual interactions when the
sequence is repeated over some specified amount of time.  This only
requires conventional decoupling sequences which are, generally,
easier to design.

The qubits are divided into four sets: idle qubits on sublattices $A$
and $B$ (depending on the chosen graph), and the sets $A'\subseteq A$
and $B'\subseteq B$ which together make up all of the pairs of
neighboring qubits where we want to preserve the couplings.  The
corresponding sequences are denoted $V_A$, $V_B$, $V_{A'}$, and
$V_{B'}$.  The universal idle-qubits sequences $V_A$ and $V_B$ must
decouple both the system (\ref{eq:ising-network}) and the bath
(\ref{eq:ising-bath}) Hamiltonians, and have sufficient flexibility so
that the coupling with a neighboring opposite-sublattice qubit driven
by the sequence $V_{B'}$ and $V_{A'}$ respectively could also be
decoupled.  On the other hand, the sequences $V_{A'}$ and $V_{B'}$
executed on the pairs of qubits to remain coupled must average out the
bath Hamiltonians (\ref{eq:ising-bath}), but leave some fraction $f$
of the original coupling (\ref{eq:ising-network}) between these
qubits.

We designed the global sequences $V_A$ and $V_B$ to allow for
construction of local versions of the sequences $V_{B'}(f)$ and
$V_{A'}(f)$, with some range of allowed fractions $f$.  This makes the
fraction $f$ locally adjustable\cite{De-Pryadko-FT-2014}, to
accommodate for possible local variations of the couplings $J_{ij}$.
In this work we assume all couplings equal (non-zero $J_{ij}=J$ iff
the sites $i$ and $j$ are connected), and use the fastest version of
these sequences of duration $\tau_\mathrm{seq}=16\tau_p$ with a fixed
fraction $f=1/2$, as used originally in
Ref.~\onlinecite{De-Pryadko-2013}.  Over the duration of the sequence,
for each pair of qubits designated to be coupled, the original
coupling $J_{ij}\equiv J$ in Eq.~(\ref{eq:ising-network}) is reduced
to $fJ=J/2$, which gives the rotation angle
$\alpha=fJ\tau_\mathrm{sec}/2$.

We constructed a {\sc cnot} gate using the
identity\cite{Galiautdinov-2007,Geller-Pritchett-Galiautdinov-Martinis-2010}
\begin{eqnarray} U_{jk}^{(\text{C-}X)} &=& Y_j X_k\bar X_j\bar Y_j
  \bar Y_k \exp\Bigl(-i\frac{\pi}{4}\sigma^z_{j}\sigma^z_k\Bigr)Y_k\\ &=& Z_j
  X_k \bar Y_k \exp\Bigl(-i\frac{\pi}{4}\sigma^z_{j}\sigma^z_{k}\Bigr)Y_k,
  \label{CNOT}
\end{eqnarray}
where, e.g., $X_k\equiv -i\sigma_k^x$ and $\bar X_k\equiv i\sigma_k^x$
respectively are the unitaries corresponding to $\pm\pi$ rotations of
the $k$-th qubit around the $X$ axis.  Eq.~(\ref{CNOT}) requires a
$ZZ$ rotation with the rotation angle $\alpha=\pi/4$.  Thus,
the pulse duration $\tau_p$ and the number of repetition
$N_\mathrm{rep}$ must satisfy the design
equation\cite{De-Pryadko-2013}
\begin{equation}
  N_\mathrm{rep}\, J \tau_p=\frac{\pi}{16}.
  \label{eq:tau_p}
\end{equation}
Larger values of $N_{\rm rep}$ correspond to smaller values of the
perturbation-theory parameter $J\tau_p$ which improves the fidelity as
it provides better decoupling.  On the other hand, this also increases
the cost in terms of the number of pulses.  The actual set of driving
fields used to implement the \textsc{cnot} gate with $N_{\rm rep}=1$
are shown in Fig.\ref{fig:cnot-seqs}.  For our calculations we used
$N_{\rm rep}=5$.

We also implemented two other controlled two-qubit gates using the
identities
\begin{eqnarray} U_{jk}^{(\text{C-}Y)} &=& e^{-i\pi/4}\bar X_2\bar Z_j
  \bar Z_k \exp\Bigl(-i\frac{\pi}{4}\sigma^z_{j}\sigma^z_{k}\Bigr)X_k
  ,
  \label{CY}\\ U_{jk}^{(\text{C-}Z)} &=& e^{-i\pi/4}\bar Z_j \bar Z_k
  \exp\Bigl(-i\frac{\pi}{4}\sigma^z_{j}\sigma^z_{k}\Bigr) , 
  \label{CZ}
\end{eqnarray}
as well as the \textsc{swap} gate as a sequence of three \textsc{cnot}
gates.\cite{Barenco-1995}

\subsection{Five-qubit code on a star graph}
\label{sec:code}
We use the smallest single-error-correcting code \cite{Bennett-1996,%
  Calderbank-Rains-Shor-Sloane-1997,Laflamme-1996} formally denoted as
$[[5,1,3]]$.  This distance-three code encodes a single qubit in a
two-dimensional subspace $\mathcal{Q}$ of the $2^5$-dimensional
Hilbert space of $n=5$ qubits.  It is a stabilizer
code\cite{gottesman-thesis}: the subspace
$$
\mathcal{Q}= \{\ket\psi:\,G_i\ket\psi=\ket\psi, \;i=1,2,\ldots r\}
$$
is a common $+1$ eigenspace of the $r=4$ independent commuting
stabilizer generators,
\begin{equation}
  \label{eq:gens}
  \begin{aligned}
    G_1= XZZXI\equiv \sigma_1^x\sigma_2^z\sigma_3^z\sigma_4^x,\\
    G_2= IXZZX\equiv\sigma_2^x\sigma_3^z\sigma_4^z\sigma_5^x,\\
    G_3= XIXZZ\equiv \sigma_1^x\sigma_3^x\sigma_4^z\sigma_5^z,\\
    G_4= ZXIXZ\equiv\sigma_1^z\sigma_2^x\sigma_4^x\sigma_5^z,
  \end{aligned}
\end{equation}
expressed as Kronecker products of single-qubit Pauli operators
$\sigma_i^\mu$, $\mu=x,y,z$.  Notice that to reduce the confusion with
the pulse unitaries in Sec.~\ref{sec:dd-seqs}, here we quote both the
commonly used positional and the traditional notations for multi-qubit
Pauli operators.

\begin{figure}[htbp]
  \raisebox{.55em}{({\bf a})}%
  \includegraphics[width=0.72\columnwidth]{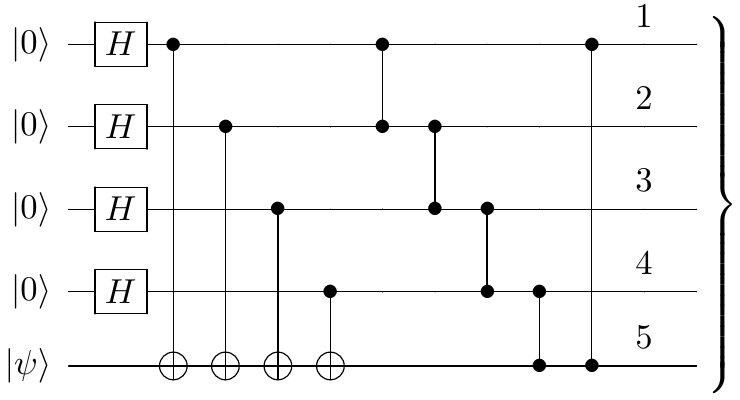}\hskip0.63in\medskip\\
  \raisebox{.25em}{({\bf b})}%
  \includegraphics[width=0.9\columnwidth]{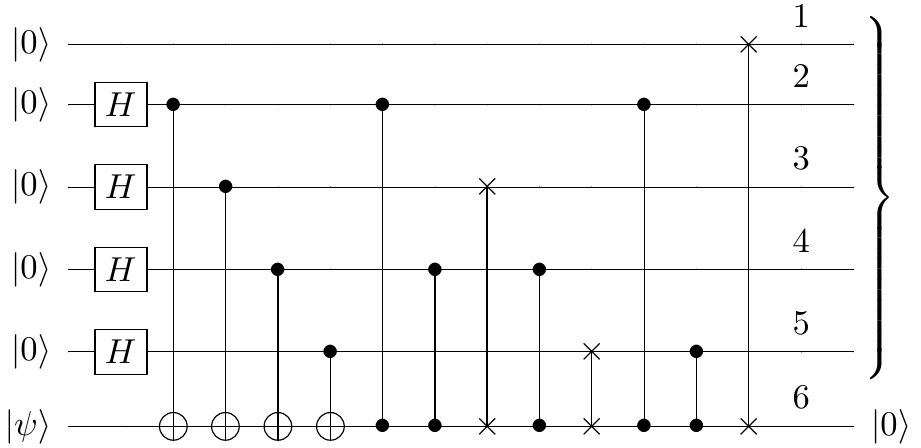}
  \caption{(\textbf{a}) Conceptual encoding circuit for the [[5,1,3]]
    code using the Hadamard, \textsc{cnot}, and controlled-phase
    gates.  On the input, the first four qubits are initialized in
    $\ket0$ states and the last qubit contains the state
    $\ket\psi\equiv\alpha\ket0+\beta\ket1$ to be encoded.  On the
    output, the corresponding logical state
    $\alpha\ket{\bar0}+\beta\ket{\bar1}$ of the five-qubit code is
    produced.  The decoding is done by inverting the encoding circuit.
    (\textbf{b}) Actual encoding circuit implemented on a six-qubit
    star graph.  Two \textsc{swap} operations are required since
    qubits on the leaves can only interact with the qubit $6$ in the
    center.  In addition, the ancilla qubit is \textsc{swap}ped to the
    center at the end to prepare for the measurement cycle, see Fig.\
    \protect\ref{fig:meas}.}
  \label{fig:encod}
\end{figure}

As for any stabilizer code, encoding of the five-qubit code can be
done efficiently\cite{gottesman-thesis}.  We have used the conceptual
encoding circuit in Fig.~\ref{fig:encod}(\textbf{a}), which produces
the code in the basis with the logical operators $\bar X=-XXXXX$ and
$\bar Z=ZZZZZ$.  This circuit is based on a representation of the
five-qubit code as a code word stabilized (CWS)
code\cite{Cross-CWS-2009}, and was constructed as a simplification of
the circuit containing the Hadamard gate on the information qubit,
encoder for the classical five-qubit repetition code, and the graph
state encoder\cite{Raussendorf-2003}.  Explicitly, the resulting basis
wavefunctions corresponding to the eigenvalues $\lambda_{\bar
  Z}=(-1)^m$, $m=0,1$, are (up to a normalization)
\begin{eqnarray}
  \nonumber
  \ket{\Psi_m}&=&
  \ket{0000m}-\ket{0110m}+\ket{1001m}-\ket{1111m}\\
  \nonumber
  &+ & \ket{0010\bar{m}}+\ket{0100\bar{m}} 
  -\ket{1101\bar{m}}-\ket{1011\bar{m}}\\
  \nonumber
  -(-)^m&\!\times&\!\!\bigl(\ket{0001\bar{m}}+
  \ket{1110\bar{m}}+\ket{0111\bar{m}}+\ket{1000\bar{m}}\\
  &+ &  \ket{0011m} -\ket{0101m}-\ket{1010m}  +\ket{1100m}\bigr).
  \qquad 
\end{eqnarray}

\begin{figure*}[htbp]
  \centering
  \includegraphics[width=0.9\textwidth]{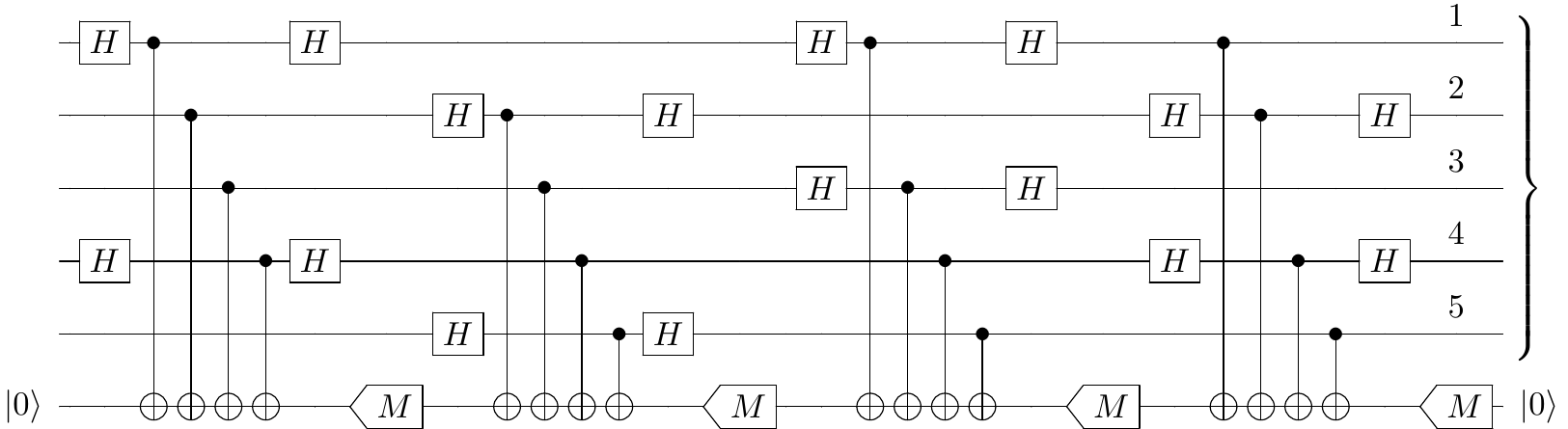}
  \caption{The circuit for a single cycle of measuring the
    stabilizer generators (\ref{eq:gens}) for the five-qubit code on a
    star graph.  On the input and the output, the first five qubits
    contain an encoded state.  The same ancilla qubit $6$ at the
    center is used for each measurement; $M$ stands for measurement
    and resetting to $\ket0$ if needed.  We implemented this circuit
    which uses 16 \textsc{cnot} gates (duration $144\tau_p$ each) and
    8 rounds of Hadamard gates (duration $32\tau_p$ each) applied in
    parallel, with the total measurement cycle duration of
    $2560\tau_p$ ($640\tau_p$ per stabilizer generator).  In our
    simulations, the entire cycle is repeated several times for
    repetitive QEC.}
  \label{fig:meas}
\end{figure*}

To implement the same circuit on the star graph, we used two more
\textsc{swap} operations, plus an additional \textsc{swap} at the end
to place the ancilla at the center, see
Fig.~\ref{fig:encod}(\textbf{b}).  This initializes for the stabilizer
generator measurement cycle shown in Fig.~\ref{fig:meas}.

\section{Simulations}
\label{sec:simulations}

We implemented the described encoding/decoding and the measurement
circuits at the Hamiltonian level, using pulse sequences described in
Sec.~\ref{sec:dd-seqs}, and classical zero-mean Gaussian phase noise
with Gaussian correlations,
\begin{equation}
  \label{eq:noise-corr}
  \langle A_i(t)\rangle=0,\quad
  \langle A_i(t)A_j(t')\rangle =\sigma^2\delta_{ij}e^{-(t-t')^2/\tau_n^2},  
\end{equation}
as a source of decoherence, cf.\ Eq.~(\ref{eq:ising-bath}).  Notice that
for a single uncontrolled qubit, such a field would produce asymptotic
dephasing rate $1/T_2=(\sqrt\pi/2)\,\sigma^2\tau_n$.

The corresponding many-body unitary dynamics has been simulated with a
{\tt C++} program using the \texttt{Eigen3} library\cite{eigenweb} for
matrix algebra.  The program uses a custom-built algorithm to schedule
the pulse sequences and measurement events, and the fourth-order
Runge-Kutta algorithm for explicitly integrating the time dependent
Schr\"odinger equation for the unitary time evolution of clusters of
multiple qubits.  In all simulations shown we used 1024 time steps per
nominal pulse duration $\tau_p$, resulting in relative integration
errors better than $10^{-9}$, comparable to numerical precision.

\subsection{Quantum error detection mode}
\label{sec:Zeno}

We first consider the working of the [[5,1,3]] code in the error
detection mode (quantum Zeno
cycle\cite{Facchi-Pascazio-2002,Facchi-2002B}).  In an actual
experiment, one is supposed to measure the stabilizer generators
repeatedly, with the experiment terminated once an error is detected
as indicated by a non-zero syndrome bit.  In our simulations, instead,
each syndrome measurement was replaced by an instantaneous projection
\begin{equation}
  \label{eq:projector}
  P_0\equiv (\openone+\sigma_6^z)/2,
\end{equation}
which selects the many-body sector with the ancilla qubit $Q_6$ at the
center in the state $\ket0$.  The success probability averaged over
the initial state $\ket\psi$ was calculated according to the
expression
\begin{equation}
  p_0\equiv \tr (U\rho_0 U^\dagger P_0), 
  \label{eq:probs}
\end{equation}
where $U$ is the $N\times N$ unitary evolution matrix up to the moment
of measurement, $\rho_0=M^{-1}P_M$ is the density matrix describing
the uniform distribution of the initial wavefunctions in a subspace of
dimensionality $M$, and $P_M$ is the corresponding projector.  In our
simulations, $N=64$ is the dimensionality of the six-qubit Hilbert
space, and we compute a reduced $N\times M$ evolution matrix $V$ which
include only $M=2$ columns corresponding to the number of basis states
of the initial qubit, see the encoding circuit in
Fig.~\ref{fig:encod}.  Respectively, we used Eq.~(\ref{eq:probs}) in
the form
\begin{equation}
  \label{eq:probs-V}
  p_0={1\over2}\tr(V^\dagger P_0 V).  
\end{equation}
Given the  reduced evolution matrix $V\equiv V(t)$ at a
given time moment $t$, and the corresponding ideal evolution matrix
$V_0$, the overall fidelity averaged over the initial
state can be calculated using the expression
\begin{equation}
  F\equiv F(V,V_0)={\tr (V_0^\dagger VV^\dagger V_0)+|\tr(V_0^\dagger
    V)|^2\over M(M+1)}.\label{eq:succ-fidelity}
\end{equation}
The derivation is similar to that given in the Appendix of
Ref.~\onlinecite{pryadko-sengupta-2008} for the case of $M=N$.
 
\begin{figure}[htbp]
  \centering
  \includegraphics[width=1\columnwidth]{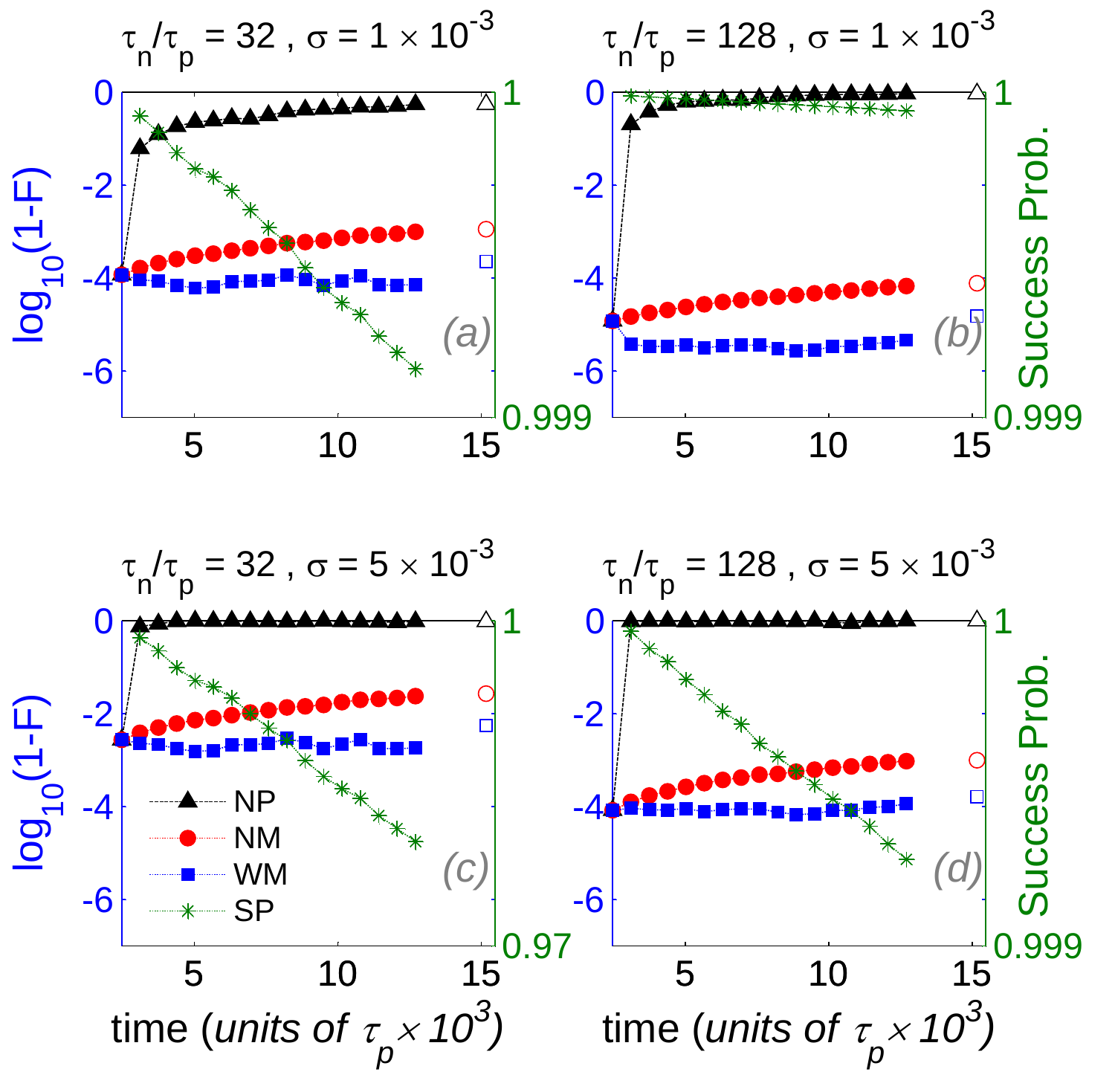}
  \caption{(color online) Infidelities during the Zeno cycle for
    different noise correlation times and noise amplitudes as
    indicated.  The time axis starts after the completion of the
    encoding circuit, at the instance of the first measurement, see
    Figs.~\ref{fig:encod}(\textbf{b}) and \ref{fig:meas}.  The
    different curves correspond to cases where no pulses are applied
    (NP), DD pulses are applied but no measurements are done (NM), and
    with the syndrome measurements (WM). Closed and open symbols
    respectively represent the infidelities at the end of each
    syndrome measurement and at the end of the final decoding. Note
    that the axis for the cumulative success probability (SP) is on
    the right.}
  \label{fig:QED}
\end{figure}

Results of simulations for several sets of parameters of the Gaussian
noise, the r.m.s.\ amplitude $\sigma$ and the correlation time
$\tau_n$, are shown in Fig.~\ref{fig:QED}.  Each plot is an average
over 20 instances of the stochastic noise, with the time axis starting
at the first measurement after the end of the encoding.  Having in
mind an experiment where the success probability and the state
fidelity would be measured separately, and to match the quantities
computed in Ref.~\onlinecite{De-Pryadko-2013}, we plot the success
probability (SP), and the fidelity ``with measurements'' (WM)
conditioned on the error-free syndrome measurements, $F_\mathrm{succ}=
F/p_0$, where $p_0$ is given by Eq.~(\ref{eq:probs-V}).  Notice that
this expression is an approximation which ignores possible
correlations between $p_0$ and $F_\mathrm{succ}$.  These correlations
would be absent with ideal syndrome measurements; we expect them to be
small in our case since the measurement fidelity is high.  The effect
of such correlations is additionally suppressed since
$F_\mathrm{succ}$ is numerically close to one.

To compare the contributions of the DD protection and of the
projective measurements (Zeno cycle) to the overall fidelity, in
Fig.~\ref{fig:QED} we also show the average fidelity
(\ref{eq:succ-fidelity}) calculated when decoupling pulses are applied
but ``no measurements'' are done (NM), and when no projective
measurements and ``no pulses'' are applied (NP).  Since they involve
no projective measurements, these quantities are independent of the
success probability (\ref{eq:probs}). For each version of the cycle,
filled symbols show the infidelities $1-F$ after each syndrome
measurement, while open symbols show the corresponding infidelities to
the end of the final decoding.  Notice that thus computed fidelities
involve all six qubits; final infidelities could be additionally
reduced by tracing out all but one information qubit, see
Sec.~\ref{sec:QEC}.

These plots show about an order of magnitude infidelity reduction due
to QEC during the cycle.  The code can detect any one- and two-qubit
error, and a small fraction of higher-weight errors.  The fact that
the Zeno cycle works, indicates that errors seen by the code are not
dominated by multi-qubit correlations.  In addition, the infidelities
increase sharply with shorter noise correlation times, as expected due
to the asymmetry of the single-qubit gates, see
Sec.~\ref{sec:dd-seqs}.

Two of the plots shown have exactly the same noise parameters and use
the same pulse shapes as in our earlier work\cite{De-Pryadko-2013}
where Zeno cycle was simulated with the $[[4,2,2]]$ error-detecting
code, with five qubits arranged in a chain.  The corresponding success
probabilities and state fidelities are similar in magnitude.  We
believe this to be a combined result of an improvement due to more
efficient code and faster syndrome measurements in the present case,
negated by increased errors due to larger connectivity of the star
graph, as discussed in detail in Ref.~\onlinecite{De-Pryadko-FT-2014}.

\subsection{QEC mode}
\label{sec:QEC}

In this mode we simulated projective measurements of the ancilla by
applying instantaneous projection operators $P_0$
[Eq.~(\ref{eq:projector})] or $P_1\equiv \openone-P_0$.  These are
six-qubit projectors selecting the sector with the ancilla qubit $Q_6$
at the center in the state $\ket0$ or $\ket1$, respectively.  Given
the normalized wavefunction $\Psi$ of the system, the projectors
should be chosen with the probabilities
$p_0\equiv\bra{\Psi}P_0\ket{\Psi}$ and
$p_1\equiv\bra{\Psi}P_1\ket{\Psi}=1-p_0$, respectively.  This implies
a separate simulation would be needed for every state $\psi$ of the
initial qubit (see the encoding circuit in Fig.~\ref{fig:encod}).
Instead, to speed up the simulations, we calculated the reduced
unitary evolution matrix $V$ and used the probability
(\ref{eq:probs-V}) averaged over the initial state of the qubit.  The
normalization of $V$ was corrected after each projection.  This
approximation is similar to that used in the previous section to
define the fidelity $F_\mathrm{succ}$ conditioned on the string of
zero-syndrome measurements in each previous cycle.  Here, we also
expect the effect of any potential unaccounted correlations to be
suppressed due to the smallness of $p_1=1-p_0$.

\begin{figure}[htbp]
  \centering
  \includegraphics[width=1\columnwidth]{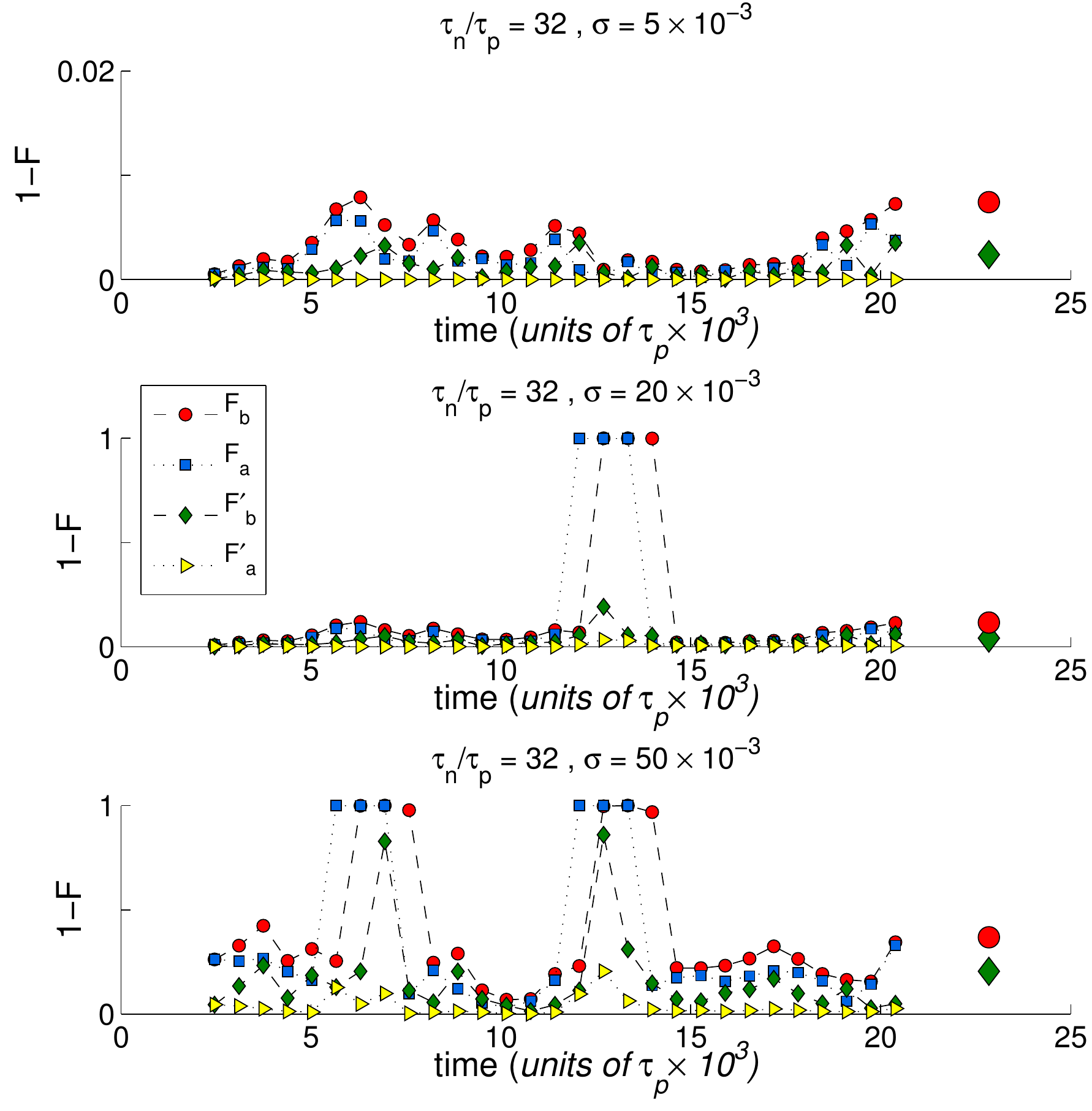}
  \caption{Sample error correction traces for $[[5,1,3]]$ code in the
    presence of the stochastic phase noise on all six qubits. The
    noise correlation time is $\tau_n=32\tau_p$ and noise amplitudes
    $\sigma$ are as indicated (in units $\tau_p^{-1}$).  Plots show
    different infidelity measures $1-F$ during the stabilizer
    measurement cycle and at the end of the decoding.  Here
    $F_\mathrm{b}$ and $F_\mathrm{a}$ are the regular fidelities
    [Eq.~(\ref{eq:succ-fidelity})], respectively, computed just before
    and right after each projective measurement.  $F'_\mathrm{b}$ and
    $F'_\mathrm{a}$ are the corresponding fidelities which include
    idealized recovery for single qubit errors, see Eq.\
    (\ref{eq:fid-recovery}).  Larger symbols indicate the infidelities
    at the end of the decoding circuit, where the circle corresponds
    to the full-system fidelity (\ref{eq:succ-fidelity}), and the
    diamond to the single-qubit fidelity after tracing out the qubits
    away from the center.}
  \label{fig:5Qerr}
\end{figure}

As in the previous section, we simulated decoherence with classical
phase noise applied on all six qubits involved in the simulations.
The noise was uncorrelated between different qubits. For each qubit,
the noise was generated as a stationary zero-mean Gaussian stochastic
process with Gaussian time correlations.  Individual traces of such
simulations for three realizations of the Gaussian stochastic noise
with identical correlation time $\tau_n=32\tau_p$ and different
r.m.s.\ amplitudes as indicated are shown in Fig.~\ref{fig:5Qerr}.
Each panel shows four different infidelity measures $1-F$ computed
during a single simulation run.  The fidelities $F_\mathrm{b}$ and
$F_\mathrm{a}$ are computed according to Eq.~(\ref{eq:succ-fidelity}),
respectively, just before and right after each projective measurement.
The fidelities $F_\mathrm{b}'$ and $F_\mathrm{a}'$ computed at the
same time moments include idealized recovery channel, see
Eq.~(\ref{eq:fid-recovery}) and the discussion below.

The five-qubit code is a ``perfect'' single-error-correcting code,
since the fifteen ($2^4-1$) non-zero syndromes corresponding to four
stabilizer generators (\ref{eq:gens}) are in a one-to-one
correspondence with the fifteen single-qubit errors.  We used this
idealized map for decoding.  Notice, however, that in our simulations
the stabilizer generators are measured sequentially, with the entire
measurement cycle typically repeated just a few times.  To increase
the syndrome measurement fidelity, we did not adhere to a fixed
measurement cycle and instead triggered the beginning of a cycle by a
syndrome measurement returning a non-zero bit.  After the fourth
measurement, the correction would be computed and applied immediately.
Typically, the infidelities $1-F_\mathrm{b}$ computed right before a
trigger event were small, whereas right after the infidelity
$1-F_\mathrm{a}$ jumps to near one, as the wavefunction is projected
outside of the code.  The infidelities $1-F_\mathrm{b}$ remain large
right before the subsequent three measurements, creating an easy to
spot four-dot pedestal in the combined trace.  For example, no trigger
events happened in the top trace in Fig.~\ref{fig:5Qerr}
($\sigma=5\times 10^{-3}/\tau_p$), one happened in the middle trace
($\sigma=20\times 10^{-3}/\tau_p$), and two in the bottom trace
($\sigma=50\times 10^{-3}/\tau_p$).

To look beyond the simple system fidelity~(\ref{eq:succ-fidelity}), we
also calculated the fidelity including the idealized recovery map,
\begin{equation}
  F'(V,V_0) = F(V,V_0)+\sum_{i=1}^{15}F(V,E_i V_0),
  \label{eq:fid-recovery}
\end{equation}
where $E_i$, $i=1,\ldots,15$, are all single-qubit errors on the
peripheral qubits, and $F(V,V_0)$ is the usual fidelity
(\ref{eq:succ-fidelity}).  This expression corresponds to idealized
error correction, with the summation over all single-qubit errors
corresponding to that over all possible syndromes.

We should mention that in our calculations both fidelity expressions
include the ancilla qubit which has not been traced out.  However,
since the ancilla is reset to $\ket0$ state after each projective
measurement, it is effectively excluded for the fidelities
$F_\mathrm{a}$ and $F_\mathrm{a}'$ computed right after the
measurement.  The ancilla is also included in the full-system fidelity
$F$ computed at the end of the decoding circuit, but not in the final
fidelity $F'$ which only looks at the state of the single qubit in the
center.  In our plots these fidelities are shown with bigger symbols.

\begin{figure}
  \centering
  \includegraphics[width=1\columnwidth]{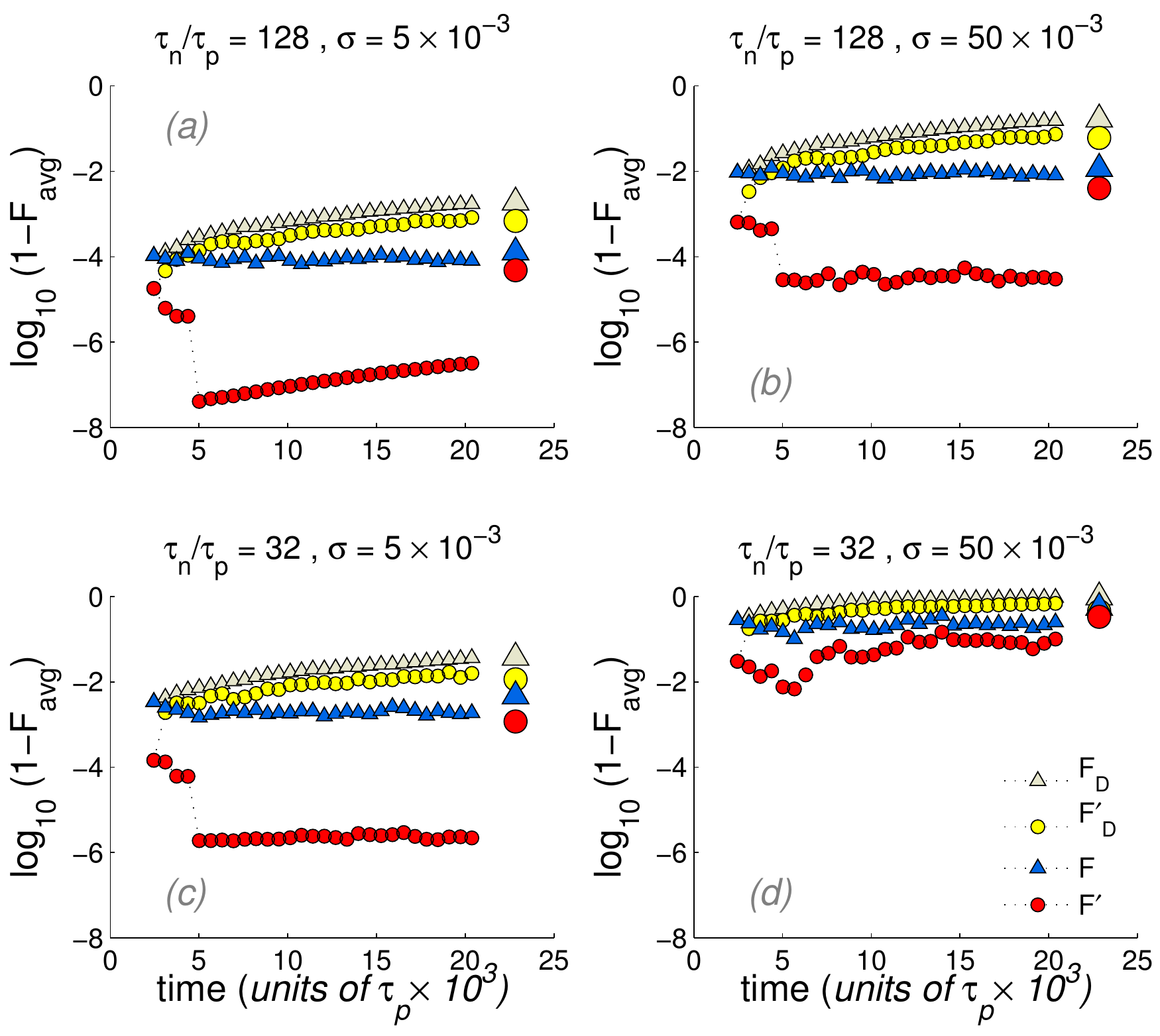}
  \caption{Infidelities $1-F$ for the $[[5,1,3]]$ code averaged over
    25 realizations of the stochastic Gaussian noise, with
    the noise correlation times $\tau_n$ and amplitudes $\sigma$ as
    indicated.  Large infidelities after trigger events have been
    excluded from the averages.  $F$ are the regular fidelities
    [Eq.~(\ref{eq:succ-fidelity})] computed right after each projective
    measurement, $F'$ are the corresponding fidelities which include
    idealized recovery for single qubit errors, see Eq.\
    (\ref{eq:fid-recovery}).  In addition $F_D$ and $F'_D$,
    respectively, are the fidelities (\ref{eq:succ-fidelity}) and
    (\ref{eq:fid-recovery}) for DD-only simulations with the same
    pulse sequences run but error correction turned off (no projective
    measurements). Larger symbols indicate the respective quantities
    at the end of the decoding circuits, with $F'$ and $F_D'$ replaced
    by the average single-qubit decoded fidelities, with the qubits
    away from the center traced out.}
  \label{fig:Qavg}
\end{figure}

The plots in Fig.~\ref{fig:5Qerr} show the simulated infidelities for
one simulation run each, they are strongly affected by the details of
the particular noise realization and the measurement results simulated
probabilistically.  In Fig.~\ref{fig:Qavg} we show (in the logarithmic
scale) infidelities averaged over 25 different realizations
of the stochastic noise.  To reduce the unphysical fluctuations, large
infidelities after the trigger events have been excluded from the
averages.

The data in Fig.~\ref{fig:Qavg} also include average infidelities
$1-F_D$ and $1-F_D'$ [Eqs.~(\ref{eq:succ-fidelity}) and
(\ref{eq:fid-recovery})] produced in identical simulations but with
error correction turned off (the same pulse sequences but no
projective measurement).  Except for the plots in
Fig.~\ref{fig:Qavg}(d), where QEC becomes relatively ineffective due
to strong noise with shorter correlation time, the DD-only
infidelities show a substantially steeper growth than those where both
DD and QEC was active.  The overall QEC effectiveness can be
quantified by comparing the final single-qubit infidelities $1-F_D'$
and $1-F'$ at the end of the decoding (two larger circles).  The
corresponding ratios of average infidelities for different panels in
Fig.~\ref{fig:Qavg} are: (a) $14.3$, (b) $15.0$, (c) $9.73$, and (d)
$1.36$.  Except for the data in Fig.~\ref{fig:Qavg}(d), QEC in these
plots gives average infidelity reduction by an order of magnitude or
better.  Notice that for this data, trigger events are rare; here QEC
fidelity is similar to that for the Zeno cycle, see
Sec.~\ref{sec:Zeno}.

In the three plots where QEC works well, Fig.~\ref{fig:Qavg}(a)--(c),
the data for $1-F'$ is some two orders of magnitude below that for
$1-F$, indicating that in the present setup single-qubit errors
strongly dominate.  This is in an apparent contrast with the results
of our Ref.~\onlinecite{De-Pryadko-FT-2014}, where we concluded that
multi-qubit errors are an unavoidable feature of the perturbatively
designed gates.  We notice, however, that due to asymmetry of
single-qubit gates, in the presence of time-dependent noise, the
leading-order error terms are single-qubit Pauli
operators\cite{De-Pryadko-FT-2014}, with the coefficients scaling as a
derivative of the classical fields $A_i(t)$.  Further correlated
errors are formed in higher orders of the Magnus series, they can be
represented as connected clusters on the connectivity graph.  On the
star graph, these include a single bond joining the ancilla at the
center to one of the code qubits, and, in the next order, two bonds,
which could result in a correlated error involving the ancilla and two
qubits of the code.  Thus, after the ancilla is projected during the
measurement, the remaining errors on the qubits forming the code are
expected to have smaller weight than they would with a different
connectivity graph.  The applicability of these arguments is improved
by our choice $N_\mathrm{rep}=5$, which gives the perturbation theory
parameter $J\tau_\mathrm{sec}=\pi/5$, where
$\tau_\mathrm{sec}=16\tau_p$ is the typical sequence duration, see
Eq.~(\ref{eq:tau_p}).

\begin{figure}[htbp]
  \centering
  \includegraphics[width=1\columnwidth]{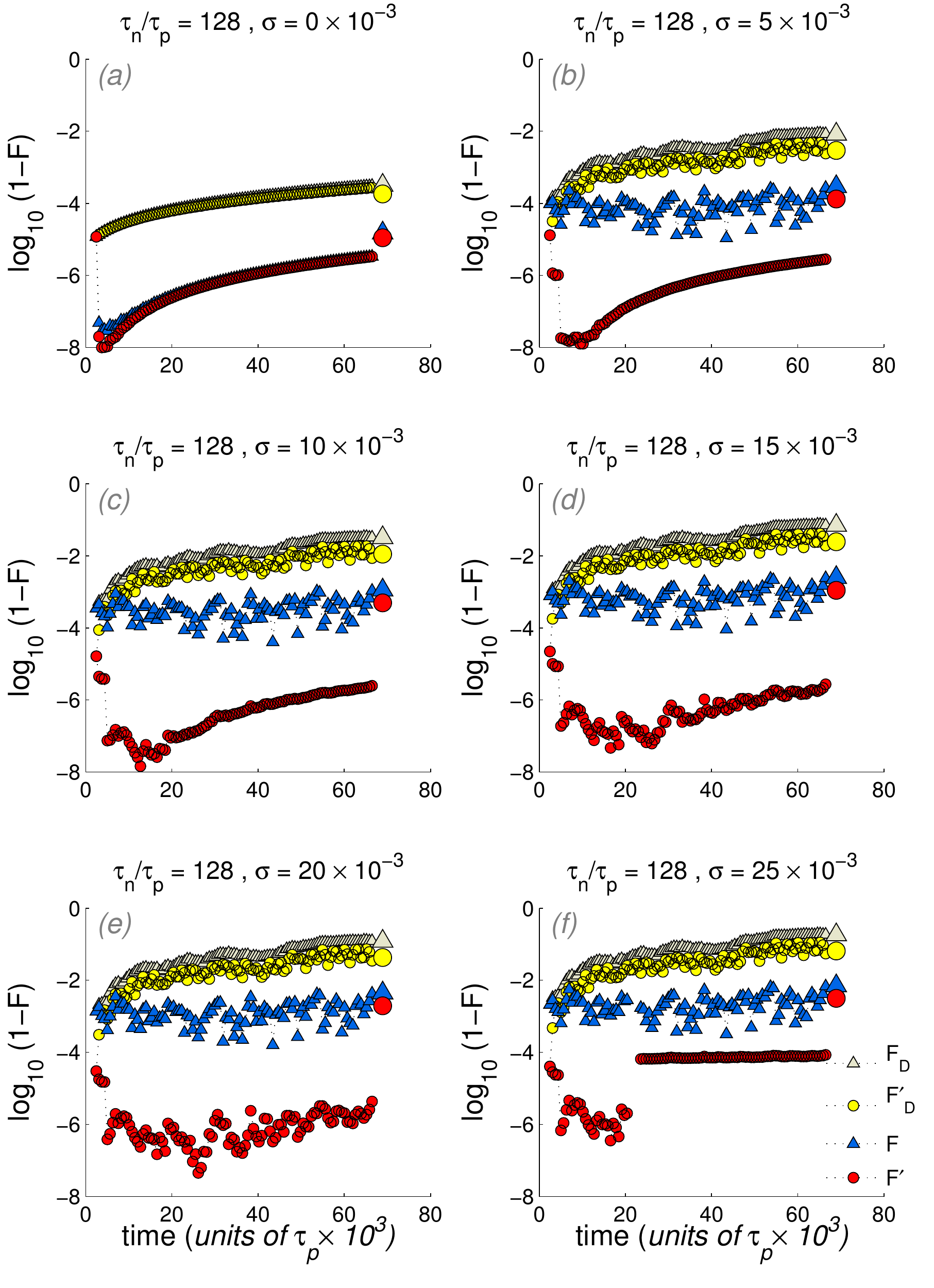}
  \caption{As in Fig.~\ref{fig:5Qerr} but with longer noise
    correlation time, $\tau_n=128\tau_p$.  Different infidelities are
    labeled as in Fig.~\ref{fig:Qavg}.}
  \label{fig:noiseprog}
\end{figure}

This analysis is confirmed by the plots in Fig.~\ref{fig:noiseprog},
which show infidelity traces for different amplitudes of the noise
with the correlation time $\tau_n=128\tau_p$.  The two top panels,
with noise amplitudes $\sigma=0$ and $\sigma=5\times 10^{-3}/\tau_p$,
show near identical plots for $F'$, indicating that with the noise
parameters as in Fig.~\ref{fig:noiseprog}(b), multi-qubit errors are
strongly dominated by the systematic errors due to the couplings $J$.
At the same time, single-qubit errors are dominated by the stochastic
noise, since at $\sigma=0$, the plots for $F$ and $F'$ fall nearly on
top of each other.

\begin{figure}
  \centering
  \includegraphics[width=1\columnwidth]{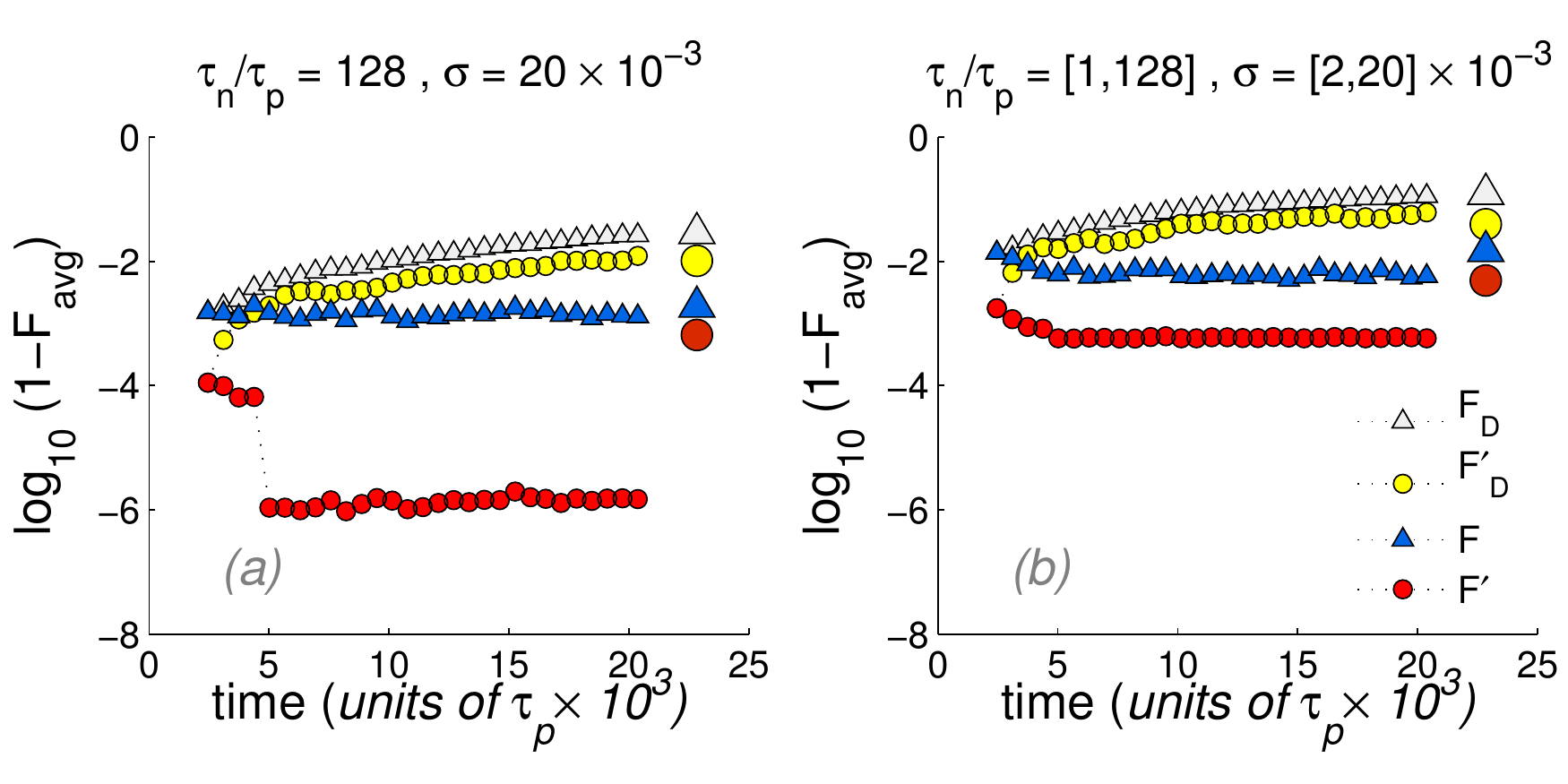}
  \caption{As in Fig.~\ref{fig:Qavg}, comparing the effect of fast
    dephasing noise.  The plots at the left show infidelities averaged 
    over 25 realizations of a Gaussian
    stochastic noise with r.m.s.\ amplitude $\sigma=20\times
    10^{-3}/\tau_p$ and $\tau_n=128\tau_p$.  For the data on the
    right, in addition, there was also a weak noise with
    $\sigma=2\times 10^{-3}/\tau_p$ and a much shorter correlation
    time $\tau_n=\tau_p$.  Such a noise is not affected by the DD.
    With the addition of the fast noise, the infidelity $1-F$
    increased by about an order of magnitude, while the infidelity
    $1-F'$ accounting by multi-qubit errors increased by two orders of
    magnitude, consistent with the expected absence of DD protection
    against the fast noise component.}
  \label{fig:2noise}
\end{figure}

Similar conclusions can be also drawn from the data in
Fig.~\ref{fig:2noise}, which shows the effect of a much faster noise,
with the correlation time $\tau_n=\tau_p$.  Namely, the infidelities
in Fig.~\ref{fig:2noise}(a) were generated by averaging the results of
25 simulations with different realizations of Gaussian noise with the
correlation time $\tau_n=128\tau_p$, while the noise for infidelities
in Fig.~\ref{fig:2noise}(b), in addition, also included weaker but
faster-varying noise components with $\tau_n=\tau_p$.  Dynamical
decoupling has nearly no effect on such a fast noise.  Respectively,
the usual infidelity $1-F$ increased by an order of magnitude, while
the infidelity $1-F'$ including idealized recovery map [see
Eq.~(\ref{eq:fid-recovery})] increased by more than two orders of
magnitude.  Such a different scaling of the two infidelities dominated
by one- and multi-qubit errors, respectively, is consistent with the
expectation of the absence of DD protection against the faster noise.
Quantitatively, the ratios of the final average single-qubit
infidelities at the end of the decoding in runs with and without error
correction are 15.7 in Fig.~\ref{fig:noiseprog}(a) and 8.0 in
Fig.~\ref{fig:noiseprog}(b).

\section{Discussion}\label{sec:conclusions}

For many years, the road to building a quantum computer appeared to be
straightforward: one just had to manufacture a sufficient number of
quality qubits and implement a universal set of quantum gates of
sufficiently high fidelity.  Now that we are there, or nearly there,
it turns out that fidelity is not the ultimate measure of performance
in large qubit systems.  As the number of qubits in a quantum computer
grows, exponentially so does the number of ways it can go wrong.  To
understand what is going on in a particular implementation of a
quantum computer would require detailed numerical simulations,
including as many qubits and as much physical detail as possible.

In this work we presented one such simulation, implementing repetitive
QEC with the $[[5,1,3]]$ code on a six-qubit network with always-on
Ising couplings and classical correlated phase noise as a source of
decoherence.  The one- and two-qubit gates were implemented via
carefully designed sequences of shaped pulses.  Realistically
simulating such gates and associated errors requires integrating the
corresponding multi-qubit unitary dynamics.  Such simulations, like
current experiments, are limited to very small system sizes.  As a
result, one can use only simplest weak codes, with very few ancillary
qubits, which puts additional constraints on the accuracy of the
implemented gates.

As in any case where gates are designed perturbatively, up to some
fixed order in the perturbation (interaction) Hamiltonian, the
systematic errors associated with our gates are correlated multi-qubit
errors, which worsen with the increased connectivity of the qubit
network.  On the other hand, the $[[5,1,3]]$ code we used is able to
correct only single-qubit errors.  To make QEC possible, we had to
tune the couplings down and make the two-qubit gates longer,
increasing the intrinsic fidelities of our gates to six nines or more.
As a result, just a few rounds of repetitive QEC required tens of
thousands of pulses, with the fidelity noticeably suffering, e.g.,
from relatively modest integration errors (not shown).  In this weakly
coupled regime, our simulations show that single-qubit errors due to
phase noise do not propagate excessively.

Overall, we demonstrated repetitive quantum error correction in a
fully quantum-mechanical simulation, with the error correction
responsible for the average infidelity reduction by an order of
magnitude or more.  We have also presented a combined protocol
integrating dynamical decoupling and quantum error correction.
Dynamical decoupling is particularly effective against low-frequency
noise which in our simulations had an asymptotic dephasing time $T_2$
as short as few nominal pulse lengths $\tau_p$.  We also
saw that our combined DD/QEC protocol remains effective in the
presence of a weak high-frequency phase noise.

While dephasing-only model appear to be too simplistic, we notice that
as a result of controlled dynamics, some of the dephasing propagates
to the longitudinal channel\cite{pryadko-quiroz-2009}.  In particular,
our original simulations which involved similar gates with the
three-qubit code protecting against single-qubit phase errors,
produced a much smaller fidelity improvement due to
QEC\cite{De-Pryadko-313-2013}.

Our model excludes many physical effects which may be relevant for
engineering a quantum computer based on a specific qubits
implementation, such as couplings between nominally disconnected
qubits, multi-level structure of the solid-state qubits and
corresponding leakage errors, violations of the rotating wave
approximation, realistic decoherence which may produce additional
correlations between the qubits, etc.  Even when the corresponding
effects are small, they can result in errors correlated in time or
between qubits, and thus strongly affect the overall coherent
multi-qubit dynamics.  Designing coherence protection schemes with
improved stability against such effects is also possible, if one knows
which decoherence mechanisms are dominant.  Each additional
improvement would require more finely tuned pulse shapes, longer
gates, or a longer code, increasing the requirements on the dynamical
range of the qubit system used in the experiment.  Thus, in our
opinion, careful studies of realistic models which incorporate such
effects are absolutely necessary in order to construct a scalable
quantum computer.

\begin{acknowledgments}
  This work was supported in part by the U.S.\ Army Research Office
  under Grant No.\ W911NF-14-1-0272 and by the NSF under Grant No.\
  PHY-1416578.  LPP also acknowledges hospitality by the Institute for
  Quantum Information and Matter, an NSF Physics Frontiers Center with
  support of the Gordon and Betty Moore Foundation.
\end{acknowledgments}

\bibliography{qc_all,lpp,more_qc,ldpc}

\begin{thebibliography}{73}%
\makeatletter
\providecommand \@ifxundefined [1]{%
 \@ifx{#1\undefined}
}%
\providecommand \@ifnum [1]{%
 \ifnum #1\expandafter \@firstoftwo
 \else \expandafter \@secondoftwo
 \fi
}%
\providecommand \@ifx [1]{%
 \ifx #1\expandafter \@firstoftwo
 \else \expandafter \@secondoftwo
 \fi
}%
\providecommand \natexlab [1]{#1}%
\providecommand \enquote  [1]{``#1''}%
\providecommand \bibnamefont  [1]{#1}%
\providecommand \bibfnamefont [1]{#1}%
\providecommand \citenamefont [1]{#1}%
\providecommand \href@noop [0]{\@secondoftwo}%
\providecommand \href [0]{\begingroup \@sanitize@url \@href}%
\providecommand \@href[1]{\@@startlink{#1}\@@href}%
\providecommand \@@href[1]{\endgroup#1\@@endlink}%
\providecommand \@sanitize@url [0]{\catcode `\\12\catcode `\$12\catcode
  `\&12\catcode `\#12\catcode `\^12\catcode `\_12\catcode `\%12\relax}%
\providecommand \@@startlink[1]{}%
\providecommand \@@endlink[0]{}%
\providecommand \url  [0]{\begingroup\@sanitize@url \@url }%
\providecommand \@url [1]{\endgroup\@href {#1}{\urlprefix }}%
\providecommand \urlprefix  [0]{URL }%
\providecommand \Eprint [0]{\href }%
\providecommand \doibase [0]{http://dx.doi.org/}%
\providecommand \selectlanguage [0]{\@gobble}%
\providecommand \bibinfo  [0]{\@secondoftwo}%
\providecommand \bibfield  [0]{\@secondoftwo}%
\providecommand \translation [1]{[#1]}%
\providecommand \BibitemOpen [0]{}%
\providecommand \bibitemStop [0]{}%
\providecommand \bibitemNoStop [0]{.\EOS\space}%
\providecommand \EOS [0]{\spacefactor3000\relax}%
\providecommand \BibitemShut  [1]{\csname bibitem#1\endcsname}%
\let\auto@bib@innerbib\@empty
\bibitem [{\citenamefont {Shor}(1995)}]{shor-error-correct}%
  \BibitemOpen
  \bibfield  {author} {\bibinfo {author} {\bibfnamefont {P.~W.}\ \bibnamefont
  {Shor}},\ }\bibfield  {title} {\enquote {\bibinfo {title} {Scheme for
  reducing decoherence in quantum computer memory},}\ }\href
  {http://link.aps.org/abstract/PRA/v52/pR2493} {\bibfield  {journal} {\bibinfo
   {journal} {Phys. Rev. A}\ }\textbf {\bibinfo {volume} {52}},\ \bibinfo
  {pages} {R2493} (\bibinfo {year} {1995})}\BibitemShut {NoStop}%
\bibitem [{\citenamefont {Gottesman}(1997)}]{gottesman-thesis}%
  \BibitemOpen
  \bibfield  {author} {\bibinfo {author} {\bibfnamefont {Daniel}\ \bibnamefont
  {Gottesman}},\ }\emph {\bibinfo {title} {Stabilizer Codes and Quantum Error
  Correction}},\ \href {http://arxiv.org/abs/quant-ph/9705052} {Ph.D. thesis},\
  \bibinfo  {school} {Caltech} (\bibinfo {year} {1997})\BibitemShut {NoStop}%
\bibitem [{\citenamefont {Knill}\ and\ \citenamefont
  {Laflamme}(1997)}]{Knill-Laflamme-1997}%
  \BibitemOpen
  \bibfield  {author} {\bibinfo {author} {\bibfnamefont {Emanuel}\ \bibnamefont
  {Knill}}\ and\ \bibinfo {author} {\bibfnamefont {Raymond}\ \bibnamefont
  {Laflamme}},\ }\bibfield  {title} {\enquote {\bibinfo {title} {Theory of
  quantum error-correcting codes},}\ }\href
  {http://dx.doi.org/10.1103/PhysRevA.55.900} {\bibfield  {journal} {\bibinfo
  {journal} {Phys. Rev. A}\ }\textbf {\bibinfo {volume} {55}},\ \bibinfo
  {pages} {900--911} (\bibinfo {year} {1997})}\BibitemShut {NoStop}%
\bibitem [{\citenamefont {Terhal}(2015)}]{Terhal-RMP-2015}%
  \BibitemOpen
  \bibfield  {author} {\bibinfo {author} {\bibfnamefont {Barbara~M.}\
  \bibnamefont {Terhal}},\ }\bibfield  {title} {\enquote {\bibinfo {title}
  {Quantum error correction for quantum memories},}\ }\href {\doibase
  10.1103/RevModPhys.87.307} {\bibfield  {journal} {\bibinfo  {journal} {Rev.
  Mod. Phys.}\ }\textbf {\bibinfo {volume} {87}},\ \bibinfo {pages} {307--346}
  (\bibinfo {year} {2015})}\BibitemShut {NoStop}%
\bibitem [{\citenamefont {Shor}(1996)}]{Shor-FT-1996}%
  \BibitemOpen
  \bibfield  {author} {\bibinfo {author} {\bibfnamefont {P.~W.}\ \bibnamefont
  {Shor}},\ }\bibfield  {title} {\enquote {\bibinfo {title} {Fault-tolerant
  quantum computation},}\ }in\ \href {http://arxiv.org/abs/quant-ph/9605011v2}
  {\emph {\bibinfo {booktitle} {Proc. 37th Ann. Symp. on Fundamentals of Comp.
  Sci.}}},\ \bibinfo {organization} {IEEE}\ (\bibinfo  {publisher} {IEEE Comp.
  Soc. Press},\ \bibinfo {address} {Los Alamitos},\ \bibinfo {year} {1996})\
  pp.\ \bibinfo {pages} {56--65},\ \Eprint
  {http://arxiv.org/abs/quant-ph/9605011} {quant-ph/9605011} \BibitemShut
  {NoStop}%
\bibitem [{\citenamefont {Steane}(1997)}]{Steane-FT-1997}%
  \BibitemOpen
  \bibfield  {author} {\bibinfo {author} {\bibfnamefont {A.~M.}\ \bibnamefont
  {Steane}},\ }\bibfield  {title} {\enquote {\bibinfo {title} {Active
  stabilization, quantum computation, and quantum state synthesis},}\ }\href
  {\doibase 10.1103/PhysRevLett.78.2252} {\bibfield  {journal} {\bibinfo
  {journal} {Phys. Rev. Lett.}\ }\textbf {\bibinfo {volume} {78}},\ \bibinfo
  {pages} {2252--2255} (\bibinfo {year} {1997})}\BibitemShut {NoStop}%
\bibitem [{\citenamefont {Gottesman}(1998)}]{Gottesman-FT-1998}%
  \BibitemOpen
  \bibfield  {author} {\bibinfo {author} {\bibfnamefont {Daniel}\ \bibnamefont
  {Gottesman}},\ }\bibfield  {title} {\enquote {\bibinfo {title} {Theory of
  fault-tolerant quantum computation},}\ }\href {\doibase
  10.1103/PhysRevA.57.127} {\bibfield  {journal} {\bibinfo  {journal} {Phys.
  Rev. A}\ }\textbf {\bibinfo {volume} {57}},\ \bibinfo {pages} {127--137}
  (\bibinfo {year} {1998})}\BibitemShut {NoStop}%
\bibitem [{\citenamefont {Dennis}\ \emph {et~al.}(2002)\citenamefont {Dennis},
  \citenamefont {Kitaev}, \citenamefont {Landahl},\ and\ \citenamefont
  {Preskill}}]{Dennis-Kitaev-Landahl-Preskill-2002}%
  \BibitemOpen
  \bibfield  {author} {\bibinfo {author} {\bibfnamefont {E.}~\bibnamefont
  {Dennis}}, \bibinfo {author} {\bibfnamefont {A.}~\bibnamefont {Kitaev}},
  \bibinfo {author} {\bibfnamefont {A.}~\bibnamefont {Landahl}}, \ and\
  \bibinfo {author} {\bibfnamefont {J.}~\bibnamefont {Preskill}},\ }\bibfield
  {title} {\enquote {\bibinfo {title} {Topological quantum memory},}\ }\href
  {http://dx.doi.org/10.1063/1.1499754} {\bibfield  {journal} {\bibinfo
  {journal} {J. Math. Phys.}\ }\textbf {\bibinfo {volume} {43}},\ \bibinfo
  {pages} {4452} (\bibinfo {year} {2002})}\BibitemShut {NoStop}%
\bibitem [{\citenamefont {Knill}(2003)}]{Knill-FT-2003}%
  \BibitemOpen
  \bibfield  {author} {\bibinfo {author} {\bibfnamefont {E.}~\bibnamefont
  {Knill}},\ }\href@noop {} {\enquote {\bibinfo {title} {Scalable quantum
  computation in the presence of large detected-error rates},}\ } (\bibinfo
  {year} {2003}),\ \bibinfo {note} {unpublished},\ \Eprint
  {http://arxiv.org/abs/arXiv:quant-ph/0312190} {arXiv:quant-ph/0312190}
  \BibitemShut {NoStop}%
\bibitem [{\citenamefont {Knill}\ \emph {et~al.}(1998)\citenamefont {Knill},
  \citenamefont {Laflamme},\ and\ \citenamefont {Zurek}}]{Knill-error-bound}%
  \BibitemOpen
  \bibfield  {author} {\bibinfo {author} {\bibfnamefont {E.}~\bibnamefont
  {Knill}}, \bibinfo {author} {\bibfnamefont {R.}~\bibnamefont {Laflamme}}, \
  and\ \bibinfo {author} {\bibfnamefont {W.~H.}\ \bibnamefont {Zurek}},\
  }\bibfield  {title} {\enquote {\bibinfo {title} {Resilient quantum
  computation},}\ }\href
  {http://www.sciencemag.org/cgi/content/abstract/279/5349/342} {\bibfield
  {journal} {\bibinfo  {journal} {Science}\ }\textbf {\bibinfo {volume}
  {279}},\ \bibinfo {pages} {342} (\bibinfo {year} {1998})}\BibitemShut
  {NoStop}%
\bibitem [{\citenamefont {Steane}(2003)}]{Steane-2003}%
  \BibitemOpen
  \bibfield  {author} {\bibinfo {author} {\bibfnamefont {Andrew~M.}\
  \bibnamefont {Steane}},\ }\bibfield  {title} {\enquote {\bibinfo {title}
  {Overhead and noise threshold of fault-tolerant quantum error correction},}\
  }\href {http://dx.doi.org/10.1103/PhysRevA.68.042322} {\bibfield  {journal}
  {\bibinfo  {journal} {Phys. Rev. A}\ }\textbf {\bibinfo {volume} {68}},\
  \bibinfo {pages} {042322} (\bibinfo {year} {2003})}\BibitemShut {NoStop}%
\bibitem [{\citenamefont {Kitaev}(2003)}]{kitaev-anyons}%
  \BibitemOpen
  \bibfield  {author} {\bibinfo {author} {\bibfnamefont {A.~Yu.}\ \bibnamefont
  {Kitaev}},\ }\bibfield  {title} {\enquote {\bibinfo {title} {Fault-tolerant
  quantum computation by anyons},}\ }\href
  {http://arxiv.org/abs/quant-ph/9707021} {\bibfield  {journal} {\bibinfo
  {journal} {Ann. Phys.}\ }\textbf {\bibinfo {volume} {303}},\ \bibinfo {pages}
  {2} (\bibinfo {year} {2003})}\BibitemShut {NoStop}%
\bibitem [{\citenamefont {Raussendorf}\ and\ \citenamefont
  {Harrington}(2007)}]{Raussendorf-Harrington-2007}%
  \BibitemOpen
  \bibfield  {author} {\bibinfo {author} {\bibfnamefont {Robert}\ \bibnamefont
  {Raussendorf}}\ and\ \bibinfo {author} {\bibfnamefont {Jim}\ \bibnamefont
  {Harrington}},\ }\bibfield  {title} {\enquote {\bibinfo {title}
  {Fault-tolerant quantum computation with high threshold in two dimensions},}\
  }\href {http://link.aps.org/abstract/PRL/v98/e190504} {\bibfield  {journal}
  {\bibinfo  {journal} {Phys. Rev. Lett.}\ }\textbf {\bibinfo {volume} {98}},\
  \bibinfo {pages} {190504} (\bibinfo {year} {2007})}\BibitemShut {NoStop}%
\bibitem [{\citenamefont {Cory}\ \emph {et~al.}(1998)\citenamefont {Cory},
  \citenamefont {Price}, \citenamefont {Maas}, \citenamefont {Knill},
  \citenamefont {Laflamme}, \citenamefont {Zurek}, \citenamefont {Havel},\ and\
  \citenamefont {Somaroo}}]{Cory-QECC-1998}%
  \BibitemOpen
  \bibfield  {author} {\bibinfo {author} {\bibfnamefont {D.~G.}\ \bibnamefont
  {Cory}}, \bibinfo {author} {\bibfnamefont {M.~D.}\ \bibnamefont {Price}},
  \bibinfo {author} {\bibfnamefont {W.}~\bibnamefont {Maas}}, \bibinfo {author}
  {\bibfnamefont {E.}~\bibnamefont {Knill}}, \bibinfo {author} {\bibfnamefont
  {R.}~\bibnamefont {Laflamme}}, \bibinfo {author} {\bibfnamefont {W.~H.}\
  \bibnamefont {Zurek}}, \bibinfo {author} {\bibfnamefont {T.~F.}\ \bibnamefont
  {Havel}}, \ and\ \bibinfo {author} {\bibfnamefont {S.~S.}\ \bibnamefont
  {Somaroo}},\ }\bibfield  {title} {\enquote {\bibinfo {title} {Experimental
  quantum error correction},}\ }\href {\doibase 10.1103/PhysRevLett.81.2152}
  {\bibfield  {journal} {\bibinfo  {journal} {Phys. Rev. Lett.}\ }\textbf
  {\bibinfo {volume} {81}},\ \bibinfo {pages} {2152--2155} (\bibinfo {year}
  {1998})}\BibitemShut {NoStop}%
\bibitem [{\citenamefont {Chiaverini}\ \emph {et~al.}(2004)\citenamefont
  {Chiaverini}, \citenamefont {Leibfried}, \citenamefont {Schaetz},
  \citenamefont {Barrett}, \citenamefont {Blakestad}, \citenamefont {Britton},
  \citenamefont {Itano}, \citenamefont {Jost}, \citenamefont {Knill},
  \citenamefont {Langer}, \citenamefont {Ozeri},\ and\ \citenamefont
  {Wineland}}]{Chiaverini-2004}%
  \BibitemOpen
  \bibfield  {author} {\bibinfo {author} {\bibfnamefont {J.}~\bibnamefont
  {Chiaverini}}, \bibinfo {author} {\bibfnamefont {D.}~\bibnamefont
  {Leibfried}}, \bibinfo {author} {\bibfnamefont {T.}~\bibnamefont {Schaetz}},
  \bibinfo {author} {\bibfnamefont {M.~D.}\ \bibnamefont {Barrett}}, \bibinfo
  {author} {\bibfnamefont {R.~B.}\ \bibnamefont {Blakestad}}, \bibinfo {author}
  {\bibfnamefont {J.}~\bibnamefont {Britton}}, \bibinfo {author} {\bibfnamefont
  {W.~M.}\ \bibnamefont {Itano}}, \bibinfo {author} {\bibfnamefont {J.~D.}\
  \bibnamefont {Jost}}, \bibinfo {author} {\bibfnamefont {E.}~\bibnamefont
  {Knill}}, \bibinfo {author} {\bibfnamefont {C.}~\bibnamefont {Langer}},
  \bibinfo {author} {\bibfnamefont {R.}~\bibnamefont {Ozeri}}, \ and\ \bibinfo
  {author} {\bibfnamefont {D.~J.}\ \bibnamefont {Wineland}},\ }\bibfield
  {title} {\enquote {\bibinfo {title} {Realization of quantum error
  correction},}\ }\href {http://dx.doi.org/10.1038/nature03074} {\bibfield
  {journal} {\bibinfo  {journal} {Nature}\ }\textbf {\bibinfo {volume} {432}},\
  \bibinfo {pages} {602} (\bibinfo {year} {2004})}\BibitemShut {NoStop}%
\bibitem [{\citenamefont {Pittman}\ \emph {et~al.}(2005)\citenamefont
  {Pittman}, \citenamefont {Jacobs},\ and\ \citenamefont
  {Franson}}]{Pittman-linear-optics-QEC-2005}%
  \BibitemOpen
  \bibfield  {author} {\bibinfo {author} {\bibfnamefont {T.~B.}\ \bibnamefont
  {Pittman}}, \bibinfo {author} {\bibfnamefont {B.~C.}\ \bibnamefont {Jacobs}},
  \ and\ \bibinfo {author} {\bibfnamefont {J.~D.}\ \bibnamefont {Franson}},\
  }\bibfield  {title} {\enquote {\bibinfo {title} {Demonstration of quantum
  error correction using linear optics},}\ }\href {\doibase
  10.1103/PhysRevA.71.052332} {\bibfield  {journal} {\bibinfo  {journal} {Phys.
  Rev. A}\ }\textbf {\bibinfo {volume} {71}},\ \bibinfo {pages} {052332}
  (\bibinfo {year} {2005})}\BibitemShut {NoStop}%
\bibitem [{\citenamefont {Schindler}\ \emph {et~al.}(2011)\citenamefont
  {Schindler}, \citenamefont {Barreiro}, \citenamefont {Monz}, \citenamefont
  {Nebendahl}, \citenamefont {Nigg}, \citenamefont {Chwalla}, \citenamefont
  {Hennrich},\ and\ \citenamefont {Blatt}}]{Schindler-Blatt-repetitive-2011}%
  \BibitemOpen
  \bibfield  {author} {\bibinfo {author} {\bibfnamefont {Philipp}\ \bibnamefont
  {Schindler}}, \bibinfo {author} {\bibfnamefont {Julio~T.}\ \bibnamefont
  {Barreiro}}, \bibinfo {author} {\bibfnamefont {Thomas}\ \bibnamefont {Monz}},
  \bibinfo {author} {\bibfnamefont {Volckmar}\ \bibnamefont {Nebendahl}},
  \bibinfo {author} {\bibfnamefont {Daniel}\ \bibnamefont {Nigg}}, \bibinfo
  {author} {\bibfnamefont {Michael}\ \bibnamefont {Chwalla}}, \bibinfo {author}
  {\bibfnamefont {Markus}\ \bibnamefont {Hennrich}}, \ and\ \bibinfo {author}
  {\bibfnamefont {Rainer}\ \bibnamefont {Blatt}},\ }\bibfield  {title}
  {\enquote {\bibinfo {title} {Experimental repetitive quantum error
  correction},}\ }\href {\doibase 10.1126/science.1203329} {\bibfield
  {journal} {\bibinfo  {journal} {Science}\ }\textbf {\bibinfo {volume}
  {332}},\ \bibinfo {pages} {1059--1061} (\bibinfo {year} {2011})},\ \Eprint
  {http://arxiv.org/abs/http://www.sciencemag.org/content/332/6033/1059.full.pdf}
  {http://www.sciencemag.org/content/332/6033/1059.full.pdf} \BibitemShut
  {NoStop}%
\bibitem [{\citenamefont {Moussa}\ \emph {et~al.}(2011)\citenamefont {Moussa},
  \citenamefont {Baugh}, \citenamefont {Ryan},\ and\ \citenamefont
  {Laflamme}}]{Moussa-NMR-QECC-2011}%
  \BibitemOpen
  \bibfield  {author} {\bibinfo {author} {\bibfnamefont {Osama}\ \bibnamefont
  {Moussa}}, \bibinfo {author} {\bibfnamefont {Jonathan}\ \bibnamefont
  {Baugh}}, \bibinfo {author} {\bibfnamefont {Colm~A.}\ \bibnamefont {Ryan}}, \
  and\ \bibinfo {author} {\bibfnamefont {Raymond}\ \bibnamefont {Laflamme}},\
  }\bibfield  {title} {\enquote {\bibinfo {title} {Demonstration of sufficient
  control for two rounds of quantum error correction in a solid state ensemble
  quantum information processor},}\ }\href {\doibase
  10.1103/PhysRevLett.107.160501} {\bibfield  {journal} {\bibinfo  {journal}
  {Phys. Rev. Lett.}\ }\textbf {\bibinfo {volume} {107}},\ \bibinfo {pages}
  {160501} (\bibinfo {year} {2011})}\BibitemShut {NoStop}%
\bibitem [{\citenamefont {Reed}\ \emph {et~al.}(2012)\citenamefont {Reed},
  \citenamefont {DiCarlo}, \citenamefont {Nigg}, \citenamefont {Sun},
  \citenamefont {Frunzio}, \citenamefont {Girvin},\ and\ \citenamefont
  {Schoelkopf}}]{Reed-QEC-SC-2012}%
  \BibitemOpen
  \bibfield  {author} {\bibinfo {author} {\bibfnamefont {M.~D.}\ \bibnamefont
  {Reed}}, \bibinfo {author} {\bibfnamefont {L.}~\bibnamefont {DiCarlo}},
  \bibinfo {author} {\bibfnamefont {S.~E.}\ \bibnamefont {Nigg}}, \bibinfo
  {author} {\bibfnamefont {L.}~\bibnamefont {Sun}}, \bibinfo {author}
  {\bibfnamefont {L.}~\bibnamefont {Frunzio}}, \bibinfo {author} {\bibfnamefont
  {S.~M.}\ \bibnamefont {Girvin}}, \ and\ \bibinfo {author} {\bibfnamefont
  {R.~J.}\ \bibnamefont {Schoelkopf}},\ }\bibfield  {title} {\enquote {\bibinfo
  {title} {Realization of three-qubit quantum error correction with
  superconducting circuits},}\ }\href {\doibase 10.1038/nature10786} {\bibfield
   {journal} {\bibinfo  {journal} {Nature}\ }\textbf {\bibinfo {volume}
  {482}},\ \bibinfo {pages} {382--385} (\bibinfo {year} {2012})}\BibitemShut
  {NoStop}%
\bibitem [{\citenamefont {Barends}\ \emph {et~al.}(2013)\citenamefont
  {Barends}, \citenamefont {Kelly}, \citenamefont {Megrant}, \citenamefont
  {Sank}, \citenamefont {Jeffrey}, \citenamefont {Chen}, \citenamefont {Yin},
  \citenamefont {Chiaro}, \citenamefont {Mutus}, \citenamefont {Neill},
  \citenamefont {O'Malley}, \citenamefont {Roushan}, \citenamefont {Wenner},
  \citenamefont {White}, \citenamefont {Cleland},\ and\ \citenamefont
  {Martinis}}]{Martinis-44us-qubit-2013}%
  \BibitemOpen
  \bibfield  {author} {\bibinfo {author} {\bibfnamefont {R.}~\bibnamefont
  {Barends}}, \bibinfo {author} {\bibfnamefont {J.}~\bibnamefont {Kelly}},
  \bibinfo {author} {\bibfnamefont {A.}~\bibnamefont {Megrant}}, \bibinfo
  {author} {\bibfnamefont {D.}~\bibnamefont {Sank}}, \bibinfo {author}
  {\bibfnamefont {E.}~\bibnamefont {Jeffrey}}, \bibinfo {author} {\bibfnamefont
  {Y.}~\bibnamefont {Chen}}, \bibinfo {author} {\bibfnamefont {Y.}~\bibnamefont
  {Yin}}, \bibinfo {author} {\bibfnamefont {B.}~\bibnamefont {Chiaro}},
  \bibinfo {author} {\bibfnamefont {J.}~\bibnamefont {Mutus}}, \bibinfo
  {author} {\bibfnamefont {C.}~\bibnamefont {Neill}}, \bibinfo {author}
  {\bibfnamefont {P.}~\bibnamefont {O'Malley}}, \bibinfo {author}
  {\bibfnamefont {P.}~\bibnamefont {Roushan}}, \bibinfo {author} {\bibfnamefont
  {J.}~\bibnamefont {Wenner}}, \bibinfo {author} {\bibfnamefont {T.~C.}\
  \bibnamefont {White}}, \bibinfo {author} {\bibfnamefont {A.~N.}\ \bibnamefont
  {Cleland}}, \ and\ \bibinfo {author} {\bibfnamefont {John~M.}\ \bibnamefont
  {Martinis}},\ }\bibfield  {title} {\enquote {\bibinfo {title} {Coherent
  josephson qubit suitable for scalable quantum integrated circuits},}\ }\href
  {\doibase 10.1103/PhysRevLett.111.080502} {\bibfield  {journal} {\bibinfo
  {journal} {Phys. Rev. Lett.}\ }\textbf {\bibinfo {volume} {111}},\ \bibinfo
  {pages} {080502} (\bibinfo {year} {2013})}\BibitemShut {NoStop}%
\bibitem [{\citenamefont {Zhong}\ \emph {et~al.}(2014)\citenamefont {Zhong},
  \citenamefont {Wang}, \citenamefont {Martinis}, \citenamefont {Cleland},
  \citenamefont {Korotkov},\ and\ \citenamefont
  {Wang}}]{Zhong-Wang-Martinis-Cleland-Korotkov-Wang-2014}%
  \BibitemOpen
  \bibfield  {author} {\bibinfo {author} {\bibfnamefont {Y.~P.}\ \bibnamefont
  {Zhong}}, \bibinfo {author} {\bibfnamefont {Z.~L.}\ \bibnamefont {Wang}},
  \bibinfo {author} {\bibfnamefont {J.~M.}\ \bibnamefont {Martinis}}, \bibinfo
  {author} {\bibfnamefont {A.~N.}\ \bibnamefont {Cleland}}, \bibinfo {author}
  {\bibfnamefont {A.~N.}\ \bibnamefont {Korotkov}}, \ and\ \bibinfo {author}
  {\bibfnamefont {H.}~\bibnamefont {Wang}},\ }\bibfield  {title} {\enquote
  {\bibinfo {title} {Reducing the impact of intrinsic dissipation in a
  superconducting circuit by quantum error detection},}\ }\href {\doibase
  10.1038/ncomms4135} {\bibfield  {journal} {\bibinfo  {journal} {Nature
  Communications}\ }\textbf {\bibinfo {volume} {5}},\ \bibinfo {pages} {3135}
  (\bibinfo {year} {2014})}\BibitemShut {NoStop}%
\bibitem [{\citenamefont {Chow}\ \emph {et~al.}(2014)\citenamefont {Chow},
  \citenamefont {Gambetta}, \citenamefont {Magesan}, \citenamefont {Abraham},
  \citenamefont {Cross}, \citenamefont {Johnson}, \citenamefont {Masluk},
  \citenamefont {Ryan}, \citenamefont {Smolin}, \citenamefont {Srinivasan},\
  and\ \citenamefont {Steffen}}]{Chow-etal-Steffen-2014}%
  \BibitemOpen
  \bibfield  {author} {\bibinfo {author} {\bibfnamefont {Jerry~M.}\
  \bibnamefont {Chow}}, \bibinfo {author} {\bibfnamefont {Jay~M.}\ \bibnamefont
  {Gambetta}}, \bibinfo {author} {\bibfnamefont {Easwar}\ \bibnamefont
  {Magesan}}, \bibinfo {author} {\bibfnamefont {David~W.}\ \bibnamefont
  {Abraham}}, \bibinfo {author} {\bibfnamefont {Andrew~W.}\ \bibnamefont
  {Cross}}, \bibinfo {author} {\bibfnamefont {B.~R.}\ \bibnamefont {Johnson}},
  \bibinfo {author} {\bibfnamefont {Nicholas~A.}\ \bibnamefont {Masluk}},
  \bibinfo {author} {\bibfnamefont {Colm~A.}\ \bibnamefont {Ryan}}, \bibinfo
  {author} {\bibfnamefont {John~A.}\ \bibnamefont {Smolin}}, \bibinfo {author}
  {\bibfnamefont {Srikanth~J.}\ \bibnamefont {Srinivasan}}, \ and\ \bibinfo
  {author} {\bibfnamefont {M.}~\bibnamefont {Steffen}},\ }\bibfield  {title}
  {\enquote {\bibinfo {title} {Implementing a strand of a scalable
  fault-tolerant quantum computing fabric},}\ }\href {\doibase
  10.1038/ncomms5015} {\bibfield  {journal} {\bibinfo  {journal} {Nature
  Communications}\ }\textbf {\bibinfo {volume} {5}},\ \bibinfo {pages} {4015}
  (\bibinfo {year} {2014})}\BibitemShut {NoStop}%
\bibitem [{\citenamefont {Barends}\ \emph {et~al.}(2014)\citenamefont
  {Barends}, \citenamefont {Kelly}, \citenamefont {Megrant}, \citenamefont
  {Veitia}, \citenamefont {Sank}, \citenamefont {Jeffrey}, \citenamefont
  {White}, \citenamefont {Mutus}, \citenamefont {Fowler}, \citenamefont
  {Campbell}, \citenamefont {Chen}, \citenamefont {Chen}, \citenamefont
  {Chiaro}, \citenamefont {Dunsworth}, \citenamefont {Neill}, \citenamefont
  {O’Malley}, \citenamefont {Roushan}, \citenamefont {Vainsencher},
  \citenamefont {Wenner}, \citenamefont {Korotkov}, \citenamefont {Cleland},\
  and\ \citenamefont {Martinis}}]{Barends-etal-Martinis-2014}%
  \BibitemOpen
  \bibfield  {author} {\bibinfo {author} {\bibfnamefont {R.}~\bibnamefont
  {Barends}}, \bibinfo {author} {\bibfnamefont {J.}~\bibnamefont {Kelly}},
  \bibinfo {author} {\bibfnamefont {A.}~\bibnamefont {Megrant}}, \bibinfo
  {author} {\bibfnamefont {A.}~\bibnamefont {Veitia}}, \bibinfo {author}
  {\bibfnamefont {D.}~\bibnamefont {Sank}}, \bibinfo {author} {\bibfnamefont
  {E.}~\bibnamefont {Jeffrey}}, \bibinfo {author} {\bibfnamefont {T.~C.}\
  \bibnamefont {White}}, \bibinfo {author} {\bibfnamefont {J.}~\bibnamefont
  {Mutus}}, \bibinfo {author} {\bibfnamefont {A.~G.}\ \bibnamefont {Fowler}},
  \bibinfo {author} {\bibfnamefont {B.}~\bibnamefont {Campbell}}, \bibinfo
  {author} {\bibfnamefont {Y.}~\bibnamefont {Chen}}, \bibinfo {author}
  {\bibfnamefont {Z.}~\bibnamefont {Chen}}, \bibinfo {author} {\bibfnamefont
  {B.}~\bibnamefont {Chiaro}}, \bibinfo {author} {\bibfnamefont
  {A.}~\bibnamefont {Dunsworth}}, \bibinfo {author} {\bibfnamefont
  {C.}~\bibnamefont {Neill}}, \bibinfo {author} {\bibfnamefont
  {P.}~\bibnamefont {O’Malley}}, \bibinfo {author} {\bibfnamefont
  {P.}~\bibnamefont {Roushan}}, \bibinfo {author} {\bibfnamefont
  {A.}~\bibnamefont {Vainsencher}}, \bibinfo {author} {\bibfnamefont
  {J.}~\bibnamefont {Wenner}}, \bibinfo {author} {\bibfnamefont {A.~N.}\
  \bibnamefont {Korotkov}}, \bibinfo {author} {\bibfnamefont {A.~N.}\
  \bibnamefont {Cleland}}, \ and\ \bibinfo {author} {\bibfnamefont {John~M.}\
  \bibnamefont {Martinis}},\ }\bibfield  {title} {\enquote {\bibinfo {title}
  {Superconducting quantum circuits at the surface code threshold for fault
  tolerance},}\ }\href {\doibase 10.1038/nature13171} {\bibfield  {journal}
  {\bibinfo  {journal} {Nature}\ }\textbf {\bibinfo {volume} {508}},\ \bibinfo
  {pages} {500--503} (\bibinfo {year} {2014})}\BibitemShut {NoStop}%
\bibitem [{\citenamefont {C{\'o}rcoles}\ \emph {et~al.}(2015)\citenamefont
  {C{\'o}rcoles}, \citenamefont {Magesan}, \citenamefont {Srinivasan},
  \citenamefont {Cross}, \citenamefont {Steffen}, \citenamefont {Gambetta},\
  and\ \citenamefont {Chow}}]{Corcoles-etal-Chow-2015}%
  \BibitemOpen
  \bibfield  {author} {\bibinfo {author} {\bibfnamefont {A.~D.}\ \bibnamefont
  {C{\'o}rcoles}}, \bibinfo {author} {\bibfnamefont {Easwar}\ \bibnamefont
  {Magesan}}, \bibinfo {author} {\bibfnamefont {Srikanth~J.}\ \bibnamefont
  {Srinivasan}}, \bibinfo {author} {\bibfnamefont {Andrew~W.}\ \bibnamefont
  {Cross}}, \bibinfo {author} {\bibfnamefont {M.}~\bibnamefont {Steffen}},
  \bibinfo {author} {\bibfnamefont {Jay~M.}\ \bibnamefont {Gambetta}}, \ and\
  \bibinfo {author} {\bibfnamefont {Jerry~M.}\ \bibnamefont {Chow}},\
  }\bibfield  {title} {\enquote {\bibinfo {title} {Demonstration of a quantum
  error detection code using a square lattice of four superconducting
  qubits},}\ }\href {\doibase 10.1038/ncomms7979} {\bibfield  {journal}
  {\bibinfo  {journal} {Nature communications}\ }\textbf {\bibinfo {volume}
  {6}} (\bibinfo {year} {2015}),\ 10.1038/ncomms7979}\BibitemShut {NoStop}%
\bibitem [{\citenamefont {Kelly}\ \emph {et~al.}(2015)\citenamefont {Kelly},
  \citenamefont {Barends}, \citenamefont {Fowler}, \citenamefont {Megrant},
  \citenamefont {Jeffrey}, \citenamefont {White}, \citenamefont {Sank},
  \citenamefont {Mutus}, \citenamefont {Campbell}, \citenamefont {Chen},
  \citenamefont {Chen}, \citenamefont {Chiaro}, \citenamefont {Dunsworth},
  \citenamefont {Hoi}, \citenamefont {Neill}, \citenamefont {O’Malley},
  \citenamefont {Quintana}, \citenamefont {Roushan}, \citenamefont
  {Vainsencher}, \citenamefont {Wenner}, \citenamefont {Cleland},\ and\
  \citenamefont {Martinis}}]{Kelly-etal-Martinis-2015}%
  \BibitemOpen
  \bibfield  {author} {\bibinfo {author} {\bibfnamefont {J.}~\bibnamefont
  {Kelly}}, \bibinfo {author} {\bibfnamefont {R.}~\bibnamefont {Barends}},
  \bibinfo {author} {\bibfnamefont {A.~G.}\ \bibnamefont {Fowler}}, \bibinfo
  {author} {\bibfnamefont {A.}~\bibnamefont {Megrant}}, \bibinfo {author}
  {\bibfnamefont {E.}~\bibnamefont {Jeffrey}}, \bibinfo {author} {\bibfnamefont
  {T.~C.}\ \bibnamefont {White}}, \bibinfo {author} {\bibfnamefont
  {D.}~\bibnamefont {Sank}}, \bibinfo {author} {\bibfnamefont {J.~Y.}\
  \bibnamefont {Mutus}}, \bibinfo {author} {\bibfnamefont {B.}~\bibnamefont
  {Campbell}}, \bibinfo {author} {\bibfnamefont {Yu}~\bibnamefont {Chen}},
  \bibinfo {author} {\bibfnamefont {Z.}~\bibnamefont {Chen}}, \bibinfo {author}
  {\bibfnamefont {B.}~\bibnamefont {Chiaro}}, \bibinfo {author} {\bibfnamefont
  {A.}~\bibnamefont {Dunsworth}}, \bibinfo {author} {\bibfnamefont {I.-C.}\
  \bibnamefont {Hoi}}, \bibinfo {author} {\bibfnamefont {C.}~\bibnamefont
  {Neill}}, \bibinfo {author} {\bibfnamefont {P.~J.~J.}\ \bibnamefont
  {O’Malley}}, \bibinfo {author} {\bibfnamefont {C.}~\bibnamefont
  {Quintana}}, \bibinfo {author} {\bibfnamefont {P.}~\bibnamefont {Roushan}},
  \bibinfo {author} {\bibfnamefont {A.}~\bibnamefont {Vainsencher}}, \bibinfo
  {author} {\bibfnamefont {J.}~\bibnamefont {Wenner}}, \bibinfo {author}
  {\bibfnamefont {A.~N.}\ \bibnamefont {Cleland}}, \ and\ \bibinfo {author}
  {\bibfnamefont {John~M.}\ \bibnamefont {Martinis}},\ }\bibfield  {title}
  {\enquote {\bibinfo {title} {State preservation by repetitive error detection
  in a superconducting quantum circuit},}\ }\href {\doibase
  10.1038/nature14270} {\bibfield  {journal} {\bibinfo  {journal} {Nature}\
  }\textbf {\bibinfo {volume} {519}},\ \bibinfo {pages} {66--69} (\bibinfo
  {year} {2015})}\BibitemShut {NoStop}%
\bibitem [{\citenamefont {Lidar}\ \emph {et~al.}(1998)\citenamefont {Lidar},
  \citenamefont {Chuang},\ and\ \citenamefont
  {Whaley}}]{Lidar-Chuang-Whaley-1998}%
  \BibitemOpen
  \bibfield  {author} {\bibinfo {author} {\bibfnamefont {D.~A.}\ \bibnamefont
  {Lidar}}, \bibinfo {author} {\bibfnamefont {I.~L.}\ \bibnamefont {Chuang}}, \
  and\ \bibinfo {author} {\bibfnamefont {K.~B.}\ \bibnamefont {Whaley}},\
  }\bibfield  {title} {\enquote {\bibinfo {title} {Decoherence-free subspaces
  for quantum computation},}\ }\href {\doibase 10.1103/PhysRevLett.81.2594}
  {\bibfield  {journal} {\bibinfo  {journal} {Phys. Rev. Lett.}\ }\textbf
  {\bibinfo {volume} {81}},\ \bibinfo {pages} {2594--2597} (\bibinfo {year}
  {1998})}\BibitemShut {NoStop}%
\bibitem [{\citenamefont {Viola}\ \emph
  {et~al.}(1999{\natexlab{a}})\citenamefont {Viola}, \citenamefont {Knill},\
  and\ \citenamefont {Lloyd}}]{viola-knill-lloyd-1999}%
  \BibitemOpen
  \bibfield  {author} {\bibinfo {author} {\bibfnamefont {Lorenza}\ \bibnamefont
  {Viola}}, \bibinfo {author} {\bibfnamefont {Emanuel}\ \bibnamefont {Knill}},
  \ and\ \bibinfo {author} {\bibfnamefont {Seth}\ \bibnamefont {Lloyd}},\
  }\bibfield  {title} {\enquote {\bibinfo {title} {Dynamical decoupling of open
  quantum systems},}\ }\href {http://link.aps.org/abstract/PRL/v82/p2417}
  {\bibfield  {journal} {\bibinfo  {journal} {Phys. Rev. Lett.}\ }\textbf
  {\bibinfo {volume} {82}},\ \bibinfo {pages} {2417} (\bibinfo {year}
  {1999}{\natexlab{a}})}\BibitemShut {NoStop}%
\bibitem [{\citenamefont {Viola}\ \emph
  {et~al.}(1999{\natexlab{b}})\citenamefont {Viola}, \citenamefont {Lloyd},\
  and\ \citenamefont {Knill}}]{viola-knill-lloyd-1999B}%
  \BibitemOpen
  \bibfield  {author} {\bibinfo {author} {\bibfnamefont {Lorenza}\ \bibnamefont
  {Viola}}, \bibinfo {author} {\bibfnamefont {Seth}\ \bibnamefont {Lloyd}}, \
  and\ \bibinfo {author} {\bibfnamefont {Emanuel}\ \bibnamefont {Knill}},\
  }\bibfield  {title} {\enquote {\bibinfo {title} {Universal control of
  decoupled quantum systems},}\ }\href
  {http://link.aps.org/abstract/PRL/v83/p4888} {\bibfield  {journal} {\bibinfo
  {journal} {Phys. Rev. Lett.}\ }\textbf {\bibinfo {volume} {83}},\ \bibinfo
  {pages} {4888} (\bibinfo {year} {1999}{\natexlab{b}})}\BibitemShut {NoStop}%
\bibitem [{\citenamefont {Lidar}\ \emph {et~al.}(1999)\citenamefont {Lidar},
  \citenamefont {Bacon},\ and\ \citenamefont {Whaley}}]{lidar-1999}%
  \BibitemOpen
  \bibfield  {author} {\bibinfo {author} {\bibfnamefont {D.~A.}\ \bibnamefont
  {Lidar}}, \bibinfo {author} {\bibfnamefont {D.}~\bibnamefont {Bacon}}, \ and\
  \bibinfo {author} {\bibfnamefont {K.~B.}\ \bibnamefont {Whaley}},\ }\bibfield
   {title} {\enquote {\bibinfo {title} {Concatenating decoherence-free
  subspaces with quantum error correcting codes},}\ }\href
  {http://link.aps.org/abstract/PRL/v82/p4556} {\bibfield  {journal} {\bibinfo
  {journal} {Phys. Rev. Lett.}\ }\textbf {\bibinfo {volume} {82}},\ \bibinfo
  {pages} {4556} (\bibinfo {year} {1999})}\BibitemShut {NoStop}%
\bibitem [{\citenamefont {Bacon}\ \emph {et~al.}(2000)\citenamefont {Bacon},
  \citenamefont {Kempe}, \citenamefont {Lidar},\ and\ \citenamefont
  {Whaley}}]{Bacon-2000}%
  \BibitemOpen
  \bibfield  {author} {\bibinfo {author} {\bibfnamefont {D.}~\bibnamefont
  {Bacon}}, \bibinfo {author} {\bibfnamefont {J.}~\bibnamefont {Kempe}},
  \bibinfo {author} {\bibfnamefont {D.~A.}\ \bibnamefont {Lidar}}, \ and\
  \bibinfo {author} {\bibfnamefont {K.~B.}\ \bibnamefont {Whaley}},\ }\bibfield
   {title} {\enquote {\bibinfo {title} {Universal fault-tolerant quantum
  computation on decoherence-free subspaces},}\ }\href
  {http://link.aps.org/abstract/PRL/v85/p1758} {\bibfield  {journal} {\bibinfo
  {journal} {Phys. Rev. Lett.}\ }\textbf {\bibinfo {volume} {85}},\ \bibinfo
  {pages} {1758} (\bibinfo {year} {2000})}\BibitemShut {NoStop}%
\bibitem [{\citenamefont {Kempe}\ \emph {et~al.}(2001)\citenamefont {Kempe},
  \citenamefont {Bacon}, \citenamefont {Lidar},\ and\ \citenamefont
  {Whaley}}]{Kempe-2001}%
  \BibitemOpen
  \bibfield  {author} {\bibinfo {author} {\bibfnamefont {J.}~\bibnamefont
  {Kempe}}, \bibinfo {author} {\bibfnamefont {D.}~\bibnamefont {Bacon}},
  \bibinfo {author} {\bibfnamefont {D.~A.}\ \bibnamefont {Lidar}}, \ and\
  \bibinfo {author} {\bibfnamefont {K.~B.}\ \bibnamefont {Whaley}},\ }\bibfield
   {title} {\enquote {\bibinfo {title} {Theory of decoherence-free
  fault-tolerant universal quantum computation},}\ }\href
  {http://link.aps.org/abstract/PRA/v63/e042307} {\bibfield  {journal}
  {\bibinfo  {journal} {Phys. Rev. A}\ }\textbf {\bibinfo {volume} {63}},\
  \bibinfo {pages} {042307} (\bibinfo {year} {2001})}\BibitemShut {NoStop}%
\bibitem [{\citenamefont {Viola}(2002)}]{Viola-2002}%
  \BibitemOpen
  \bibfield  {author} {\bibinfo {author} {\bibfnamefont {Lorenza}\ \bibnamefont
  {Viola}},\ }\bibfield  {title} {\enquote {\bibinfo {title} {Quantum control
  via encoded dynamical decoupling},}\ }\href
  {http://dx.doi.org/10.1103/PhysRevA.66.012307} {\bibfield  {journal}
  {\bibinfo  {journal} {Phys. Rev. A}\ }\textbf {\bibinfo {volume} {66}},\
  \bibinfo {pages} {012307} (\bibinfo {year} {2002})}\BibitemShut {NoStop}%
\bibitem [{\citenamefont {Facchi}\ \emph {et~al.}(2005)\citenamefont {Facchi},
  \citenamefont {Tasaki}, \citenamefont {Pascazio}, \citenamefont {Nakazato},
  \citenamefont {Tokuse},\ and\ \citenamefont {Lidar}}]{facchi-nakazato-2004}%
  \BibitemOpen
  \bibfield  {author} {\bibinfo {author} {\bibfnamefont {P.}~\bibnamefont
  {Facchi}}, \bibinfo {author} {\bibfnamefont {S.}~\bibnamefont {Tasaki}},
  \bibinfo {author} {\bibfnamefont {S.}~\bibnamefont {Pascazio}}, \bibinfo
  {author} {\bibfnamefont {H.}~\bibnamefont {Nakazato}}, \bibinfo {author}
  {\bibfnamefont {A.}~\bibnamefont {Tokuse}}, \ and\ \bibinfo {author}
  {\bibfnamefont {D.~A.}\ \bibnamefont {Lidar}},\ }\bibfield  {title} {\enquote
  {\bibinfo {title} {Control of decoherence: Analysis and comparison of three
  different strategies},}\ }\href
  {http://link.aps.org/abstract/PRA/v71/e022302} {\bibfield  {journal}
  {\bibinfo  {journal} {Phys. Rev. A}\ }\textbf {\bibinfo {volume} {71}},\
  \bibinfo {pages} {022302} (\bibinfo {year} {2005})}\BibitemShut {NoStop}%
\bibitem [{\citenamefont {Lidar}(2014)}]{Lidar-review-2014}%
  \BibitemOpen
  \bibfield  {author} {\bibinfo {author} {\bibfnamefont {Daniel~A.}\
  \bibnamefont {Lidar}},\ }\bibfield  {title} {\enquote {\bibinfo {title}
  {Review of decoherence-free subspaces, noiseless subsystems, and dynamical
  decoupling},}\ }in\ \href {\doibase 10.1002/9781118742631.ch11} {\emph
  {\bibinfo {booktitle} {Quantum Information and Computation for Chemistry}}},\
  \bibinfo {series and number} {Advances in Chemical Physics},\ \bibinfo
  {editor} {edited by\ \bibinfo {editor} {\bibfnamefont {Sabre}\ \bibnamefont
  {Kais}}}\ (\bibinfo  {publisher} {John Wiley {\&} Sons, Inc.},\ \bibinfo
  {year} {2014})\ Chap.~\bibinfo {chapter} {11}, pp.\ \bibinfo {pages}
  {295--354}\BibitemShut {NoStop}%
\bibitem [{\citenamefont {Shiokawa}\ and\ \citenamefont
  {Lidar}(2004)}]{shiokawa-lidar-2004}%
  \BibitemOpen
  \bibfield  {author} {\bibinfo {author} {\bibfnamefont {K.}~\bibnamefont
  {Shiokawa}}\ and\ \bibinfo {author} {\bibfnamefont {D.~A.}\ \bibnamefont
  {Lidar}},\ }\bibfield  {title} {\enquote {\bibinfo {title} {Dynamical
  decoupling using slow pulses: Efficient suppression of {$1/f$} noise},}\
  }\href {http://link.aps.org/abstract/PRA/v69/e030302} {\bibfield  {journal}
  {\bibinfo  {journal} {Phys. Rev. A}\ }\textbf {\bibinfo {volume} {69}},\
  \bibinfo {pages} {030302(R)} (\bibinfo {year} {2004})}\BibitemShut {NoStop}%
\bibitem [{\citenamefont {Faoro}\ and\ \citenamefont
  {Viola}(2004)}]{faoro-viola-2004}%
  \BibitemOpen
  \bibfield  {author} {\bibinfo {author} {\bibfnamefont {Lara}\ \bibnamefont
  {Faoro}}\ and\ \bibinfo {author} {\bibfnamefont {Lorenza}\ \bibnamefont
  {Viola}},\ }\bibfield  {title} {\enquote {\bibinfo {title} {Dynamical
  suppression of {$1/f$} noise processes in qubit systems},}\ }\href
  {http://link.aps.org/abstract/PRL/v92/e117905} {\bibfield  {journal}
  {\bibinfo  {journal} {Phys. Rev. Lett.}\ }\textbf {\bibinfo {volume} {92}},\
  \bibinfo {pages} {117905} (\bibinfo {year} {2004})}\BibitemShut {NoStop}%
\bibitem [{\citenamefont {Sengupta}\ and\ \citenamefont
  {Pryadko}(2005)}]{sengupta-pryadko-ref-2005}%
  \BibitemOpen
  \bibfield  {author} {\bibinfo {author} {\bibfnamefont {P.}~\bibnamefont
  {Sengupta}}\ and\ \bibinfo {author} {\bibfnamefont {L.~P.}\ \bibnamefont
  {Pryadko}},\ }\bibfield  {title} {\enquote {\bibinfo {title} {Scalable design
  of tailored soft pulses for coherent control},}\ }\href
  {http://link.aps.org/abstract/PRL/v95/e037202} {\bibfield  {journal}
  {\bibinfo  {journal} {Phys. Rev. Lett.}\ }\textbf {\bibinfo {volume} {95}},\
  \bibinfo {pages} {037202} (\bibinfo {year} {2005})}\BibitemShut {NoStop}%
\bibitem [{\citenamefont {Pryadko}\ and\ \citenamefont
  {Sengupta}(2006)}]{pryadko-sengupta-kinetics-2006}%
  \BibitemOpen
  \bibfield  {author} {\bibinfo {author} {\bibfnamefont {L.~P.}\ \bibnamefont
  {Pryadko}}\ and\ \bibinfo {author} {\bibfnamefont {P.}~\bibnamefont
  {Sengupta}},\ }\bibfield  {title} {\enquote {\bibinfo {title} {Quantum
  kinetics of an open system in the presence of periodic refocusing fields},}\
  }\href {http://link.aps.org/abstract/PRB/v73/e085321} {\bibfield  {journal}
  {\bibinfo  {journal} {Phys. Rev. B}\ }\textbf {\bibinfo {volume} {73}},\
  \bibinfo {pages} {085321} (\bibinfo {year} {2006})}\BibitemShut {NoStop}%
\bibitem [{\citenamefont {Kuopanportti}\ \emph {et~al.}(2008)\citenamefont
  {Kuopanportti}, \citenamefont {M\"ott\"onen}, \citenamefont {Bergholm},
  \citenamefont {Saira}, \citenamefont {Zhang},\ and\ \citenamefont
  {Whaley}}]{kuopanportti-2008}%
  \BibitemOpen
  \bibfield  {author} {\bibinfo {author} {\bibfnamefont {Pekko}\ \bibnamefont
  {Kuopanportti}}, \bibinfo {author} {\bibfnamefont {Mikko}\ \bibnamefont
  {M\"ott\"onen}}, \bibinfo {author} {\bibfnamefont {Ville}\ \bibnamefont
  {Bergholm}}, \bibinfo {author} {\bibfnamefont {Olli-Pentti}\ \bibnamefont
  {Saira}}, \bibinfo {author} {\bibfnamefont {Jun}\ \bibnamefont {Zhang}}, \
  and\ \bibinfo {author} {\bibfnamefont {K.~Birgitta}\ \bibnamefont {Whaley}},\
  }\bibfield  {title} {\enquote {\bibinfo {title} {Suppression of 1/falpha
  noise in one-qubit systems},}\ }\href
  {http://link.aps.org/abstract/PRA/v77/e032334} {\bibfield  {journal}
  {\bibinfo  {journal} {Phys. Rev. A}\ }\textbf {\bibinfo {volume} {77}},\
  \bibinfo {pages} {032334} (\bibinfo {year} {2008})}\BibitemShut {NoStop}%
\bibitem [{\citenamefont {Cywi\'nski}\ \emph {et~al.}(2008)\citenamefont
  {Cywi\'nski}, \citenamefont {Lutchyn}, \citenamefont {Nave},\ and\
  \citenamefont {Sarma}}]{Cywinski-2008}%
  \BibitemOpen
  \bibfield  {author} {\bibinfo {author} {\bibfnamefont {Lukasz}\ \bibnamefont
  {Cywi\'nski}}, \bibinfo {author} {\bibfnamefont {Roman~M.}\ \bibnamefont
  {Lutchyn}}, \bibinfo {author} {\bibfnamefont {Cody~P.}\ \bibnamefont {Nave}},
  \ and\ \bibinfo {author} {\bibfnamefont {S.~Das}\ \bibnamefont {Sarma}},\
  }\bibfield  {title} {\enquote {\bibinfo {title} {How to enhance dephasing
  time in superconducting qubits},}\ }\href
  {http://link.aps.org/doi/10.1103/PhysRevB.77.174509} {\bibfield  {journal}
  {\bibinfo  {journal} {Phys. Rev. B}\ }\textbf {\bibinfo {volume} {77}},\
  \bibinfo {pages} {174509} (\bibinfo {year} {2008})}\BibitemShut {NoStop}%
\bibitem [{\citenamefont {West}\ \emph {et~al.}(2010)\citenamefont {West},
  \citenamefont {Lidar}, \citenamefont {Fong},\ and\ \citenamefont
  {Gyure}}]{West-Lidar-Fong-Gyure-2010}%
  \BibitemOpen
  \bibfield  {author} {\bibinfo {author} {\bibfnamefont {Jacob~R.}\
  \bibnamefont {West}}, \bibinfo {author} {\bibfnamefont {Daniel~A.}\
  \bibnamefont {Lidar}}, \bibinfo {author} {\bibfnamefont {Bryan~H.}\
  \bibnamefont {Fong}}, \ and\ \bibinfo {author} {\bibfnamefont {Mark~F.}\
  \bibnamefont {Gyure}},\ }\bibfield  {title} {\enquote {\bibinfo {title} {High
  fidelity quantum gates via dynamical decoupling},}\ }\href {\doibase
  10.1103/PhysRevLett.105.230503} {\bibfield  {journal} {\bibinfo  {journal}
  {Phys. Rev. Lett.}\ }\textbf {\bibinfo {volume} {105}},\ \bibinfo {pages}
  {230503} (\bibinfo {year} {2010})}\BibitemShut {NoStop}%
\bibitem [{\citenamefont {Slichter}(1992)}]{slichter-book}%
  \BibitemOpen
  \bibfield  {author} {\bibinfo {author} {\bibfnamefont {C.~P.}\ \bibnamefont
  {Slichter}},\ }\href@noop {} {\emph {\bibinfo {title} {Principles of Magnetic
  Resonance}}},\ \bibinfo {edition} {3rd}\ ed.\ (\bibinfo  {publisher}
  {Springer-Verlag},\ \bibinfo {address} {New York},\ \bibinfo {year}
  {1992})\BibitemShut {NoStop}%
\bibitem [{\citenamefont {Stollsteimer}\ and\ \citenamefont
  {Mahler}(2001)}]{stollsteimer-mahler-2001}%
  \BibitemOpen
  \bibfield  {author} {\bibinfo {author} {\bibfnamefont {Marcus}\ \bibnamefont
  {Stollsteimer}}\ and\ \bibinfo {author} {\bibfnamefont {G{\"u}nter}\
  \bibnamefont {Mahler}},\ }\bibfield  {title} {\enquote {\bibinfo {title}
  {Suppression of arbitrary internal coupling in a quantum register},}\ }\href
  {http://link.aps.org/abstract/PRA/v64/e052301} {\bibfield  {journal}
  {\bibinfo  {journal} {Phys. Rev. A}\ }\textbf {\bibinfo {volume} {64}},\
  \bibinfo {pages} {052301} (\bibinfo {year} {2001})}\BibitemShut {NoStop}%
\bibitem [{\citenamefont {Tomita}\ \emph {et~al.}(2010)\citenamefont {Tomita},
  \citenamefont {Merrill},\ and\ \citenamefont {Brown}}]{Tomita-2010}%
  \BibitemOpen
  \bibfield  {author} {\bibinfo {author} {\bibfnamefont {Y.}~\bibnamefont
  {Tomita}}, \bibinfo {author} {\bibfnamefont {J.~T.}\ \bibnamefont {Merrill}},
  \ and\ \bibinfo {author} {\bibfnamefont {K.~R.}\ \bibnamefont {Brown}},\
  }\bibfield  {title} {\enquote {\bibinfo {title} {Multi-qubit compensation
  sequences},}\ }\href {http://stacks.iop.org/1367-2630/12/i=1/a=015002}
  {\bibfield  {journal} {\bibinfo  {journal} {New J. Phys.}\ }\textbf {\bibinfo
  {volume} {12}},\ \bibinfo {pages} {015002} (\bibinfo {year}
  {2010})}\BibitemShut {NoStop}%
\bibitem [{\citenamefont {Pryadko}\ and\ \citenamefont
  {Quiroz}(2007)}]{pryadko-quiroz-2007}%
  \BibitemOpen
  \bibfield  {author} {\bibinfo {author} {\bibfnamefont {L.~P.}\ \bibnamefont
  {Pryadko}}\ and\ \bibinfo {author} {\bibfnamefont {G.}~\bibnamefont
  {Quiroz}},\ }\bibfield  {title} {\enquote {\bibinfo {title} {Soft-pulse
  dynamical decoupling in a cavity},}\ }\href
  {http://link.aps.org/abstract/PRA/v77/e012330} {\bibfield  {journal}
  {\bibinfo  {journal} {Phys. Rev. A}\ }\textbf {\bibinfo {volume} {77}},\
  \bibinfo {pages} {012330/1--9} (\bibinfo {year} {2007})}\BibitemShut
  {NoStop}%
\bibitem [{\citenamefont {Pryadko}\ and\ \citenamefont
  {Quiroz}(2009)}]{pryadko-quiroz-2009}%
  \BibitemOpen
  \bibfield  {author} {\bibinfo {author} {\bibfnamefont {L.~P.}\ \bibnamefont
  {Pryadko}}\ and\ \bibinfo {author} {\bibfnamefont {Gregory}\ \bibnamefont
  {Quiroz}},\ }\bibfield  {title} {\enquote {\bibinfo {title} {Soft-pulse
  dynamical decoupling with {M}arkovian decoherence},}\ }\href
  {http://link.aps.org/abstract/PRA/v80/e042317} {\bibfield  {journal}
  {\bibinfo  {journal} {Phys. Rev. A}\ }\textbf {\bibinfo {volume} {80}},\
  \bibinfo {pages} {042317} (\bibinfo {year} {2009})}\BibitemShut {NoStop}%
\bibitem [{\citenamefont {Pryadko}\ and\ \citenamefont
  {Sengupta}(2008)}]{pryadko-sengupta-2008}%
  \BibitemOpen
  \bibfield  {author} {\bibinfo {author} {\bibfnamefont {L.~P.}\ \bibnamefont
  {Pryadko}}\ and\ \bibinfo {author} {\bibfnamefont {P.}~\bibnamefont
  {Sengupta}},\ }\bibfield  {title} {\enquote {\bibinfo {title} {Second-order
  shaped pulses for solid-state quantum computation},}\ }\href {\doibase
  10.1103/PhysRevA.78.032336} {\bibfield  {journal} {\bibinfo  {journal} {Phys.
  Rev. A}\ }\textbf {\bibinfo {volume} {78}},\ \bibinfo {pages} {032336}
  (\bibinfo {year} {2008})}\BibitemShut {NoStop}%
\bibitem [{\citenamefont {Kabytayev}\ \emph {et~al.}(2014)\citenamefont
  {Kabytayev}, \citenamefont {Green}, \citenamefont {Khodjasteh}, \citenamefont
  {Biercuk}, \citenamefont {Viola},\ and\ \citenamefont
  {Brown}}]{Kabytayev-Green-Khodjasteh-Biercuk-Viola-Brown-2014}%
  \BibitemOpen
  \bibfield  {author} {\bibinfo {author} {\bibfnamefont {Chingiz}\ \bibnamefont
  {Kabytayev}}, \bibinfo {author} {\bibfnamefont {Todd~J.}\ \bibnamefont
  {Green}}, \bibinfo {author} {\bibfnamefont {Kaveh}\ \bibnamefont
  {Khodjasteh}}, \bibinfo {author} {\bibfnamefont {Michael~J.}\ \bibnamefont
  {Biercuk}}, \bibinfo {author} {\bibfnamefont {Lorenza}\ \bibnamefont
  {Viola}}, \ and\ \bibinfo {author} {\bibfnamefont {Kenneth~R.}\ \bibnamefont
  {Brown}},\ }\bibfield  {title} {\enquote {\bibinfo {title} {Robustness of
  composite pulses to time-dependent control noise},}\ }\href {\doibase
  10.1103/PhysRevA.90.012316} {\bibfield  {journal} {\bibinfo  {journal} {Phys.
  Rev. A}\ }\textbf {\bibinfo {volume} {90}},\ \bibinfo {pages} {012316}
  (\bibinfo {year} {2014})}\BibitemShut {NoStop}%
\bibitem [{\citenamefont {De}\ and\ \citenamefont
  {Pryadko}(2013{\natexlab{a}})}]{De-Pryadko-2013}%
  \BibitemOpen
  \bibfield  {author} {\bibinfo {author} {\bibfnamefont {A.}~\bibnamefont
  {De}}\ and\ \bibinfo {author} {\bibfnamefont {L.~P.}\ \bibnamefont
  {Pryadko}},\ }\bibfield  {title} {\enquote {\bibinfo {title} {Universal set
  of scalable dynamically corrected gates for quantum error correction with
  always-on qubit couplings},}\ }\href {\doibase
  10.1103/PhysRevLett.110.070503} {\bibfield  {journal} {\bibinfo  {journal}
  {Phys. Rev. Lett.}\ }\textbf {\bibinfo {volume} {110}},\ \bibinfo {pages}
  {070503} (\bibinfo {year} {2013}{\natexlab{a}})},\ \Eprint
  {http://arxiv.org/abs/arXiv:1209.2764} {arXiv:1209.2764} \BibitemShut
  {NoStop}%
\bibitem [{\citenamefont {De}\ and\ \citenamefont
  {Pryadko}(2014)}]{De-Pryadko-FT-2014}%
  \BibitemOpen
  \bibfield  {author} {\bibinfo {author} {\bibfnamefont {Amrit}\ \bibnamefont
  {De}}\ and\ \bibinfo {author} {\bibfnamefont {Leonid~P.}\ \bibnamefont
  {Pryadko}},\ }\bibfield  {title} {\enquote {\bibinfo {title} {Dynamically
  corrected gates for qubits with always-on ising couplings: Error model and
  fault tolerance with the toric code},}\ }\href {\doibase
  10.1103/PhysRevA.89.032332} {\bibfield  {journal} {\bibinfo  {journal} {Phys.
  Rev. A}\ }\textbf {\bibinfo {volume} {89}},\ \bibinfo {pages} {032332}
  (\bibinfo {year} {2014})},\ \Eprint {http://arxiv.org/abs/1310.1652}
  {1310.1652} \BibitemShut {NoStop}%
\bibitem [{Note1()}]{Note1}%
  \BibitemOpen
  \bibinfo {note} {Note that this is exactly the arrangement chosen for
  experiments in Ref.\ \protect \rev@citealp
  {Barends-etal-Martinis-2014}.}\BibitemShut {Stop}%
\bibitem [{\citenamefont {Kovalev}\ \emph {et~al.}(2011)\citenamefont
  {Kovalev}, \citenamefont {Dumer},\ and\ \citenamefont
  {Pryadko}}]{Kovalev-Dumer-Pryadko-2011}%
  \BibitemOpen
  \bibfield  {author} {\bibinfo {author} {\bibfnamefont {A.~A.}\ \bibnamefont
  {Kovalev}}, \bibinfo {author} {\bibfnamefont {I.}~\bibnamefont {Dumer}}, \
  and\ \bibinfo {author} {\bibfnamefont {L.~P.}\ \bibnamefont {Pryadko}},\
  }\bibfield  {title} {\enquote {\bibinfo {title} {Design of additive quantum
  codes via the code-word-stabilized framework},}\ }\href {\doibase
  10.1103/PhysRevA.84.062319} {\bibfield  {journal} {\bibinfo  {journal} {Phys.
  Rev. A}\ }\textbf {\bibinfo {volume} {84}},\ \bibinfo {pages} {062319}
  (\bibinfo {year} {2011})}\BibitemShut {NoStop}%
\bibitem [{\citenamefont {Kovalev}\ and\ \citenamefont
  {Pryadko}(2012)}]{Kovalev-Pryadko-2012}%
  \BibitemOpen
  \bibfield  {author} {\bibinfo {author} {\bibfnamefont {A.~A.}\ \bibnamefont
  {Kovalev}}\ and\ \bibinfo {author} {\bibfnamefont {L.~P.}\ \bibnamefont
  {Pryadko}},\ }\bibfield  {title} {\enquote {\bibinfo {title} {Improved
  quantum hypergraph-product {LDPC} codes},}\ }in\ \href {\doibase
  10.1109/ISIT.2012.6284206} {\emph {\bibinfo {booktitle} {Proc. IEEE Int.
  Symp. Inf. Theory (ISIT)}}}\ (\bibinfo {year} {2012})\ pp.\ \bibinfo {pages}
  {348--352},\ \Eprint {http://arxiv.org/abs/arXiv:1202.0928} {arXiv:1202.0928}
  \BibitemShut {NoStop}%
\bibitem [{Note2()}]{Note2}%
  \BibitemOpen
  \bibinfo {note} {Selective decoupling sequences for more general qubit
  interaction Hamiltonians have been constructed, e.g., in Refs.~\protect
  \rev@citealp
  {sengupta-pryadko-ref-2005,Frydrych-Marthaler-Alber-2015}.}\BibitemShut
  {Stop}%
\bibitem [{\citenamefont {Frydrych}\ \emph {et~al.}(2015)\citenamefont
  {Frydrych}, \citenamefont {Marthaler},\ and\ \citenamefont
  {Alber}}]{Frydrych-Marthaler-Alber-2015}%
  \BibitemOpen
  \bibfield  {author} {\bibinfo {author} {\bibfnamefont {Holger}\ \bibnamefont
  {Frydrych}}, \bibinfo {author} {\bibfnamefont {Michael}\ \bibnamefont
  {Marthaler}}, \ and\ \bibinfo {author} {\bibfnamefont {Gernot}\ \bibnamefont
  {Alber}},\ }\href {http://arxiv.org/abs/1502.03665} {\enquote {\bibinfo
  {title} {Pulse-controlled quantum gate sequences on a strongly coupled qubit
  chain},}\ } (\bibinfo {year} {2015}),\ \bibinfo {note} {unpublished},\
  \Eprint {http://arxiv.org/abs/1502.03665} {1502.03665} \BibitemShut {NoStop}%
\bibitem [{\citenamefont {Warren}(1984)}]{warren-herm}%
  \BibitemOpen
  \bibfield  {author} {\bibinfo {author} {\bibfnamefont {W.~S.}\ \bibnamefont
  {Warren}},\ }\bibfield  {title} {\enquote {\bibinfo {title} {Effects of
  arbitrary laser or nmr pulse shapes on population inversion and coherence},}\
  }\href {http://dx.doi.org/10.1063/1.447644} {\bibfield  {journal} {\bibinfo
  {journal} {J. Chem. Phys.}\ }\textbf {\bibinfo {volume} {81}},\ \bibinfo
  {pages} {5437--5448} (\bibinfo {year} {1984})}\BibitemShut {NoStop}%
\bibitem [{\citenamefont {Barenco}\ \emph {et~al.}(1995)\citenamefont
  {Barenco}, \citenamefont {Bennett}, \citenamefont {Cleve}, \citenamefont
  {DiVincenzo}, \citenamefont {Margolus}, \citenamefont {Shor}, \citenamefont
  {Sleator}, \citenamefont {Smolin},\ and\ \citenamefont
  {Weinfurter}}]{Barenco-1995}%
  \BibitemOpen
  \bibfield  {author} {\bibinfo {author} {\bibfnamefont {Adriano}\ \bibnamefont
  {Barenco}}, \bibinfo {author} {\bibfnamefont {Charles~H.}\ \bibnamefont
  {Bennett}}, \bibinfo {author} {\bibfnamefont {Richard}\ \bibnamefont
  {Cleve}}, \bibinfo {author} {\bibfnamefont {David~P.}\ \bibnamefont
  {DiVincenzo}}, \bibinfo {author} {\bibfnamefont {Norman}\ \bibnamefont
  {Margolus}}, \bibinfo {author} {\bibfnamefont {Peter}\ \bibnamefont {Shor}},
  \bibinfo {author} {\bibfnamefont {Tycho}\ \bibnamefont {Sleator}}, \bibinfo
  {author} {\bibfnamefont {John~A.}\ \bibnamefont {Smolin}}, \ and\ \bibinfo
  {author} {\bibfnamefont {Harald}\ \bibnamefont {Weinfurter}},\ }\bibfield
  {title} {\enquote {\bibinfo {title} {Elementary gates for quantum
  computation},}\ }\href {http://dx.doi.org/10.1103/PhysRevA.52.3457}
  {\bibfield  {journal} {\bibinfo  {journal} {Phys. Rev. A}\ }\textbf {\bibinfo
  {volume} {52}},\ \bibinfo {pages} {3457--3467} (\bibinfo {year}
  {1995})}\BibitemShut {NoStop}%
\bibitem [{\citenamefont {Khodjasteh}\ and\ \citenamefont
  {Viola}(2009{\natexlab{a}})}]{Khodjasteh-Viola-PRL-2009}%
  \BibitemOpen
  \bibfield  {author} {\bibinfo {author} {\bibfnamefont {Kaveh}\ \bibnamefont
  {Khodjasteh}}\ and\ \bibinfo {author} {\bibfnamefont {Lorenza}\ \bibnamefont
  {Viola}},\ }\bibfield  {title} {\enquote {\bibinfo {title} {Dynamically
  error-corrected gates for universal quantum computation},}\ }\href
  {http://link.aps.org/abstract/PRL/v102/e080501} {\bibfield  {journal}
  {\bibinfo  {journal} {Phys. Rev. Lett.}\ }\textbf {\bibinfo {volume} {102}},\
  \bibinfo {pages} {080501} (\bibinfo {year} {2009}{\natexlab{a}})}\BibitemShut
  {NoStop}%
\bibitem [{\citenamefont {Khodjasteh}\ and\ \citenamefont
  {Viola}(2009{\natexlab{b}})}]{Khodjasteh-Viola-PRA-2009}%
  \BibitemOpen
  \bibfield  {author} {\bibinfo {author} {\bibfnamefont {Kaveh}\ \bibnamefont
  {Khodjasteh}}\ and\ \bibinfo {author} {\bibfnamefont {Lorenza}\ \bibnamefont
  {Viola}},\ }\bibfield  {title} {\enquote {\bibinfo {title} {Dynamical quantum
  error correction of unitary operations with bounded controls},}\ }\href
  {http://link.aps.org/abstract/PRA/v80/e032314} {\bibfield  {journal}
  {\bibinfo  {journal} {Phys. Rev. A}\ }\textbf {\bibinfo {volume} {80}},\
  \bibinfo {pages} {032314} (\bibinfo {year} {2009}{\natexlab{b}})}\BibitemShut
  {NoStop}%
\bibitem [{\citenamefont {Viola}\ and\ \citenamefont
  {Knill}(2003)}]{Viola-Knill-2003}%
  \BibitemOpen
  \bibfield  {author} {\bibinfo {author} {\bibfnamefont {Lorenza}\ \bibnamefont
  {Viola}}\ and\ \bibinfo {author} {\bibfnamefont {Emanuel}\ \bibnamefont
  {Knill}},\ }\bibfield  {title} {\enquote {\bibinfo {title} {Robust dynamical
  decoupling of quantum systems with bounded controls},}\ }\href
  {http://dx.doi.org/10.1103/PhysRevLett.90.037901} {\bibfield  {journal}
  {\bibinfo  {journal} {Phys. Rev. Lett.}\ }\textbf {\bibinfo {volume} {90}},\
  \bibinfo {pages} {037901} (\bibinfo {year} {2003})}\BibitemShut {NoStop}%
\bibitem [{\citenamefont {Pasini}\ \emph {et~al.}(2008)\citenamefont {Pasini},
  \citenamefont {Fischer}, \citenamefont {Karbach},\ and\ \citenamefont
  {Uhrig}}]{Pasini-2008}%
  \BibitemOpen
  \bibfield  {author} {\bibinfo {author} {\bibfnamefont {S.}~\bibnamefont
  {Pasini}}, \bibinfo {author} {\bibfnamefont {T.}~\bibnamefont {Fischer}},
  \bibinfo {author} {\bibfnamefont {P.}~\bibnamefont {Karbach}}, \ and\
  \bibinfo {author} {\bibfnamefont {G.~S.}\ \bibnamefont {Uhrig}},\ }\bibfield
  {title} {\enquote {\bibinfo {title} {Optimization of short coherent control
  pulses},}\ }\href {http://link.aps.org/doi/10.1103/PhysRevA.77.032315}
  {\bibfield  {journal} {\bibinfo  {journal} {Phys. Rev. A}\ }\textbf {\bibinfo
  {volume} {77}},\ \bibinfo {pages} {032315} (\bibinfo {year}
  {2008})}\BibitemShut {NoStop}%
\bibitem [{\citenamefont {Khodjasteh}\ \emph {et~al.}(2010)\citenamefont
  {Khodjasteh}, \citenamefont {Lidar},\ and\ \citenamefont
  {Viola}}]{Khodjasteh-Lidar-Viola-2009}%
  \BibitemOpen
  \bibfield  {author} {\bibinfo {author} {\bibfnamefont {Kaveh}\ \bibnamefont
  {Khodjasteh}}, \bibinfo {author} {\bibfnamefont {Daniel~A.}\ \bibnamefont
  {Lidar}}, \ and\ \bibinfo {author} {\bibfnamefont {Lorenza}\ \bibnamefont
  {Viola}},\ }\bibfield  {title} {\enquote {\bibinfo {title} {Arbitrarily
  accurate dynamical control in open quantum systems},}\ }\href {\doibase
  10.1103/PhysRevLett.104.090501} {\bibfield  {journal} {\bibinfo  {journal}
  {Phys. Rev. Lett.}\ }\textbf {\bibinfo {volume} {104}},\ \bibinfo {pages}
  {090501} (\bibinfo {year} {2010})}\BibitemShut {NoStop}%
\bibitem [{\citenamefont {Galiautdinov}(2007)}]{Galiautdinov-2007}%
  \BibitemOpen
  \bibfield  {author} {\bibinfo {author} {\bibfnamefont {Andrei}\ \bibnamefont
  {Galiautdinov}},\ }\bibfield  {title} {\enquote {\bibinfo {title} {Generation
  of high-fidelity controlled-{NOT} logic gates by coupled superconducting
  qubits},}\ }\href {\doibase 10.1103/PhysRevA.75.052303} {\bibfield  {journal}
  {\bibinfo  {journal} {Phys. Rev. A}\ }\textbf {\bibinfo {volume} {75}},\
  \bibinfo {pages} {052303} (\bibinfo {year} {2007})}\BibitemShut {NoStop}%
\bibitem [{\citenamefont {Geller}\ \emph {et~al.}(2010)\citenamefont {Geller},
  \citenamefont {Pritchett}, \citenamefont {Galiautdinov},\ and\ \citenamefont
  {Martinis}}]{Geller-Pritchett-Galiautdinov-Martinis-2010}%
  \BibitemOpen
  \bibfield  {author} {\bibinfo {author} {\bibfnamefont {Michael~R.}\
  \bibnamefont {Geller}}, \bibinfo {author} {\bibfnamefont {Emily~J.}\
  \bibnamefont {Pritchett}}, \bibinfo {author} {\bibfnamefont {Andrei}\
  \bibnamefont {Galiautdinov}}, \ and\ \bibinfo {author} {\bibfnamefont
  {John~M.}\ \bibnamefont {Martinis}},\ }\bibfield  {title} {\enquote {\bibinfo
  {title} {Quantum logic with weakly coupled qubits},}\ }\href {\doibase
  10.1103/PhysRevA.81.012320} {\bibfield  {journal} {\bibinfo  {journal} {Phys.
  Rev. A}\ }\textbf {\bibinfo {volume} {81}},\ \bibinfo {pages} {012320}
  (\bibinfo {year} {2010})}\BibitemShut {NoStop}%
\bibitem [{\citenamefont {Bennett}\ \emph {et~al.}(1996)\citenamefont
  {Bennett}, \citenamefont {DiVincenzo}, \citenamefont {Smolin},\ and\
  \citenamefont {Wootters}}]{Bennett-1996}%
  \BibitemOpen
  \bibfield  {author} {\bibinfo {author} {\bibfnamefont {C.}~\bibnamefont
  {Bennett}}, \bibinfo {author} {\bibfnamefont {D.}~\bibnamefont {DiVincenzo}},
  \bibinfo {author} {\bibfnamefont {J.}~\bibnamefont {Smolin}}, \ and\ \bibinfo
  {author} {\bibfnamefont {W.}~\bibnamefont {Wootters}},\ }\bibfield  {title}
  {\enquote {\bibinfo {title} {Mixed state entanglement and quantum error
  correction},}\ }\href {http://dx.doi.org/10.1103/PhysRevA.54.3824} {\bibfield
   {journal} {\bibinfo  {journal} {Phys. Rev. A}\ }\textbf {\bibinfo {volume}
  {54}},\ \bibinfo {pages} {3824} (\bibinfo {year} {1996})}\BibitemShut
  {NoStop}%
\bibitem [{\citenamefont {Calderbank}\ \emph {et~al.}(1997)\citenamefont
  {Calderbank}, \citenamefont {Rains}, \citenamefont {Shor},\ and\
  \citenamefont {Sloane}}]{Calderbank-Rains-Shor-Sloane-1997}%
  \BibitemOpen
  \bibfield  {author} {\bibinfo {author} {\bibfnamefont {A.~R.}\ \bibnamefont
  {Calderbank}}, \bibinfo {author} {\bibfnamefont {E.~M.}\ \bibnamefont
  {Rains}}, \bibinfo {author} {\bibfnamefont {P.~W.}\ \bibnamefont {Shor}}, \
  and\ \bibinfo {author} {\bibfnamefont {N.~J.~A.}\ \bibnamefont {Sloane}},\
  }\bibfield  {title} {\enquote {\bibinfo {title} {Quantum error correction and
  orthogonal geometry},}\ }\href {http://dx.doi.org/10.1103/PhysRevLett.78.405}
  {\bibfield  {journal} {\bibinfo  {journal} {Phys. Rev. Lett.}\ }\textbf
  {\bibinfo {volume} {78}},\ \bibinfo {pages} {405--408} (\bibinfo {year}
  {1997})}\BibitemShut {NoStop}%
\bibitem [{\citenamefont {Laflamme}\ \emph {et~al.}(1996)\citenamefont
  {Laflamme}, \citenamefont {Miquel}, \citenamefont {Paz},\ and\ \citenamefont
  {Zurek}}]{Laflamme-1996}%
  \BibitemOpen
  \bibfield  {author} {\bibinfo {author} {\bibfnamefont {Raymond}\ \bibnamefont
  {Laflamme}}, \bibinfo {author} {\bibfnamefont {Cesar}\ \bibnamefont
  {Miquel}}, \bibinfo {author} {\bibfnamefont {Juan~Pablo}\ \bibnamefont
  {Paz}}, \ and\ \bibinfo {author} {\bibfnamefont {Wojciech~Hubert}\
  \bibnamefont {Zurek}},\ }\bibfield  {title} {\enquote {\bibinfo {title}
  {Perfect quantum error correcting code},}\ }\href {\doibase
  10.1103/PhysRevLett.77.198} {\bibfield  {journal} {\bibinfo  {journal} {Phys.
  Rev. Lett.}\ }\textbf {\bibinfo {volume} {77}},\ \bibinfo {pages} {198--201}
  (\bibinfo {year} {1996})}\BibitemShut {NoStop}%
\bibitem [{\citenamefont {Cross}\ \emph {et~al.}(2009)\citenamefont {Cross},
  \citenamefont {Smith}, \citenamefont {Smolin},\ and\ \citenamefont
  {Zeng}}]{Cross-CWS-2009}%
  \BibitemOpen
  \bibfield  {author} {\bibinfo {author} {\bibfnamefont {A.}~\bibnamefont
  {Cross}}, \bibinfo {author} {\bibfnamefont {G.}~\bibnamefont {Smith}},
  \bibinfo {author} {\bibfnamefont {J.~A.}\ \bibnamefont {Smolin}}, \ and\
  \bibinfo {author} {\bibfnamefont {Bei}\ \bibnamefont {Zeng}},\ }\bibfield
  {title} {\enquote {\bibinfo {title} {Codeword stabilized quantum codes},}\
  }\href {http://dx.doi.org/10.1109/TIT.2008.2008136} {\bibfield  {journal}
  {\bibinfo  {journal} {IEEE Trans. Info. Th.}\ }\textbf {\bibinfo {volume}
  {55}},\ \bibinfo {pages} {433--438} (\bibinfo {year} {2009})}\BibitemShut
  {NoStop}%
\bibitem [{\citenamefont {Raussendorf}\ \emph {et~al.}(2003)\citenamefont
  {Raussendorf}, \citenamefont {Browne},\ and\ \citenamefont
  {Briegel}}]{Raussendorf-2003}%
  \BibitemOpen
  \bibfield  {author} {\bibinfo {author} {\bibfnamefont {Robert}\ \bibnamefont
  {Raussendorf}}, \bibinfo {author} {\bibfnamefont {Daniel~E.}\ \bibnamefont
  {Browne}}, \ and\ \bibinfo {author} {\bibfnamefont {Hans~J.}\ \bibnamefont
  {Briegel}},\ }\bibfield  {title} {\enquote {\bibinfo {title}
  {Measurement-based quantum computation on cluster states},}\ }\href
  {http://dx.doi.org/10.1103/PhysRevA.68.022312} {\bibfield  {journal}
  {\bibinfo  {journal} {Phys. Rev. A}\ }\textbf {\bibinfo {volume} {68}},\
  \bibinfo {pages} {022312} (\bibinfo {year} {2003})}\BibitemShut {NoStop}%
\bibitem [{\citenamefont {Guennebaud}\ \emph {et~al.}(2010)\citenamefont
  {Guennebaud}, \citenamefont {Jacob} \emph {et~al.}}]{eigenweb}%
  \BibitemOpen
  \bibfield  {author} {\bibinfo {author} {\bibfnamefont {Ga\"{e}l}\
  \bibnamefont {Guennebaud}}, \bibinfo {author} {\bibfnamefont {Beno\^{i}t}\
  \bibnamefont {Jacob}},  \emph {et~al.},\ }\href@noop {} {\enquote {\bibinfo
  {title} {Eigen v3},}\ }\bibinfo {howpublished} {http://eigen.tuxfamily.org}
  (\bibinfo {year} {2010})\BibitemShut {NoStop}%
\bibitem [{\citenamefont {Facchi}\ and\ \citenamefont
  {Pascazio}(2002)}]{Facchi-Pascazio-2002}%
  \BibitemOpen
  \bibfield  {author} {\bibinfo {author} {\bibfnamefont {P.}~\bibnamefont
  {Facchi}}\ and\ \bibinfo {author} {\bibfnamefont {S.}~\bibnamefont
  {Pascazio}},\ }\bibfield  {title} {\enquote {\bibinfo {title} {Quantum zeno
  subspaces},}\ }\href {http://link.aps.org/abstract/PRL/v89/e080401}
  {\bibfield  {journal} {\bibinfo  {journal} {Phys. Rev. Lett.}\ }\textbf
  {\bibinfo {volume} {89}},\ \bibinfo {pages} {080401} (\bibinfo {year}
  {2002})}\BibitemShut {NoStop}%
\bibitem [{\citenamefont {Facchi}\ \emph {et~al.}(2002)\citenamefont {Facchi},
  \citenamefont {Pascazio}, \citenamefont {Scardicchio},\ and\ \citenamefont
  {Schulman}}]{Facchi-2002B}%
  \BibitemOpen
  \bibfield  {author} {\bibinfo {author} {\bibfnamefont {P.}~\bibnamefont
  {Facchi}}, \bibinfo {author} {\bibfnamefont {S.}~\bibnamefont {Pascazio}},
  \bibinfo {author} {\bibfnamefont {A.}~\bibnamefont {Scardicchio}}, \ and\
  \bibinfo {author} {\bibfnamefont {L.~S.}\ \bibnamefont {Schulman}},\
  }\bibfield  {title} {\enquote {\bibinfo {title} {Zeno dynamics yields
  ordinary constraints},}\ }\href
  {http://link.aps.org/abstract/PRA/v65/e012108} {\bibfield  {journal}
  {\bibinfo  {journal} {Phys. Rev. A}\ }\textbf {\bibinfo {volume} {65}},\
  \bibinfo {pages} {012108} (\bibinfo {year} {2002})}\BibitemShut {NoStop}%
\bibitem [{\citenamefont {De}\ and\ \citenamefont
  {Pryadko}(2013{\natexlab{b}})}]{De-Pryadko-313-2013}%
  \BibitemOpen
  \bibfield  {author} {\bibinfo {author} {\bibfnamefont {Amrit}\ \bibnamefont
  {De}}\ and\ \bibinfo {author} {\bibfnamefont {Leonid~P.}\ \bibnamefont
  {Pryadko}},\ }\href@noop {} {\enquote {\bibinfo {title} {Simulations of the
  three-qubit code},}\ } (\bibinfo {year} {2013}{\natexlab{b}}),\ \bibinfo
  {note} {unpublished}\BibitemShut {NoStop}%
\end{thebibliography}%

\end{document}